%% file: CAGE.tex
\documentclass[sigconf]{acmart}
\AtBeginDocument{%
  \providecommand\BibTeX{{%
    \normalfont B\kern-0.5em{\scshape i\kern-0.25em b}\kern-0.8em\TeX}}}

\copyrightyear{2024}
\acmYear{2024}
\setcopyright{acmlicensed}\acmConference[WWW '24]{Proceedings of the ACM
Web Conference 2024}{May 13--17, 2024}{Singapore, Singapore}
\acmBooktitle{Proceedings of the ACM Web Conference 2024 (WWW '24), May
13--17, 2024, Singapore, Singapore}
\acmDOI{10.1145/3589334.3645484}
\acmISBN{979-8-4007-0171-9/24/05}

\acmPrice{15.00}

\acmSubmissionID{1037}

\usepackage{multirow}
\usepackage{subcaption}
\usepackage{booktabs}
\usepackage{amsfonts}
\usepackage{amsmath}
\usepackage{makecell}

\usepackage{pgfplots}
\pgfplotsset{width=7.5cm,compat=1.12}
\usepgfplotslibrary{fillbetween}

\usepackage{pgfkeys}
    \newenvironment{customlegend}[1][]{%
        \begingroup
        \csname pgfplots@init@cleared@structures\endcsname
        \pgfplotsset{#1}%
    }{%
        \csname pgfplots@createlegend\endcsname
        \endgroup
    }%
    \def\addlegendimage{\csname pgfplots@addlegendimage\endcsname}

\newcommand{\model}{CAGE}

\begin{document}

\title[CAGE]{Learning Category Trees for ID-Based Recommendation: Exploring the Power of Differentiable Vector Quantization}

\author{Qijiong Liu}
\affiliation{%
  \institution{The HK PolyU}
  \city{Hong Kong SAR}
  \country{China}
}
\email{liu@qijiong.work}

\author{Lu Fan}
\authornote{Both authors contributed equally to this research (co-second authors).}
\affiliation{%
  \institution{The HK PolyU}
  \city{Hong Kong SAR}
  \country{China}
}
\email{cslfan@comp.polyu.edu.hk}

\author{Jiaren Xiao}
\authornotemark[1]
\affiliation{%
  \institution{The HK PolyU}
  \city{Hong Kong SAR}
  \country{China}
}
\email{jiaren.xiao@polyu.edu.hk}

\author{Jieming Zhu}
\affiliation{%
  \institution{Huawei Noah's Ark Lab}
  \city{Shenzhen}
  \country{China}
}
\email{jiemingzhu@ieee.org}

\author{Xiao-Ming Wu}
\authornote{Corresponding author.}
\affiliation{%
  \institution{The HK PolyU}
  \city{Hong Kong SAR}
  \country{China}
}
\email{xiao-ming.wu@polyu.edu.hk}

\renewcommand{\shortauthors}{Qijiong Liu, Jiaren Xiao, Lu Fan, Jieming Zhu, \& Xiao-Ming Wu}

\input{content/Abstract}





\maketitle

\input{content/Introduction}
\input{content/RelatedWork}

\input{content/Method}

\input{content/Application}
\input{content/Experiment}
\input{content/Conclusion}

\bibliographystyle{ACM-Reference-Format}
\bibliography{CAGE}

\appendix
\input{content/Appendix}

\end{document}

%% file: content/Abstract.tex
\begin{abstract}

Category information plays a crucial role in enhancing the quality and personalization of recommender systems. Nevertheless, the availability of item category information is not consistently present, particularly in the context of ID-based recommendations. In this work, we propose a novel approach to automatically learn and generate entity (i.e., user or item) category trees for ID-based recommendation. Specifically, we devise a differentiable vector quantization framework for automatic category tree generation, namely CAGE, which enables the simultaneous learning and refinement of categorical code representations and entity embeddings in an end-to-end manner, starting from the randomly initialized states. With its high adaptability, CAGE can be easily integrated into both sequential and non-sequential recommender systems. We validate the effectiveness of CAGE on various recommendation tasks including list completion, collaborative filtering, and click-through rate prediction, across different recommendation models. We release the code and data\footnote{\url{https://github.com/Jyonn/Cage/}} for others to reproduce the reported results.
\end{abstract}

%% file: content/Introduction.tex
\section{Introduction}

\input{figures/dove}

Recommender systems~\cite{dcn,lightgcn,bert4rec} aim to ease the burden of decision-making by automatically suggesting personalized item recommendations tailored to a user's preferences and historical behavior. They cater to diverse objectives such as list completion, collaborative filtering, and click-through rate prediction. The varied objectives underscore the importance of devising methodologies that can adapt to different recommendation scenarios and deliver improved recommendations.

When crafting recommendation models and algorithms, the integration of categorical information is of paramount importance.
Categorical attributes, such as product types~\cite{cai2021category} and user locations~\cite{liu2022prec,moreira2019importance}, find widespread use due to their ability to capture crucial attributes and establish meaningful connections for users and items. 
% Consequently, recommender systems can learn diverse granularities of entities (e.g., user or item) representations. 
Furthermore, these category features serve to mitigate the cold-start problem, providing an additional layer of information for less active (sparsely interacting) entities (i.e., users or items)~\cite{gogna2015comprehensive,barman2019genre}. This supplementary information is progressively refined by interactions from active users or items during training, thereby aiding less active entities in obtaining more robust representations.

%\textbf{Our Work.}
However, category features are not always available, since many recommendation datasets only have ID information. To address the absence of category attributes in ID-based recommendation contexts, we propose a novel automatic \textbf{ca}tegory tree \textbf{ge}neration framework, namely \model{}, under the assumption that hierarchical categorical structures commonly arise in recommendation scenarios of varying scales. As illustrated in~\autoref{fig:illustration}, \model{} serves as the precursor to the recommender system, dynamically constructing an item/user category tree, which incorporates hierarchical categorical knowledge (e.g., ``Good'' and ``Gaming'') relevant to the current entity (e.g., ``I-123''). The categorical information, encoded as vectors, is provided to the recommender system as auxiliary information alongside the user/item ID. The implementation of \model{} is based on differentiable and cascaded vector quantization (VQ). 
%Specifically, we employ vector quantization (VQ) techniques as the core clustering algorithm. However, a direct use of VQ poses two challenges, as detailed below.
% automatically generate categorical features. However, the application of VQ poses two challenges, as detailed below. 

Previous vector quantization methods for recommendation~\cite{zhang2023query,rajput2023recommender} often rely on \textit{meaningful and fixed} entity (e.g., user or item) embeddings, derived from side information like content-aware item embeddings using pretrained models. They commonly adopt a three-stage design, as displayed in~\autoref{fig:vq-compare}, where representation learning, vector quantization, and recommendation training are carried out separately. However, the lack of side information in ID-based recommendation hinders the generation of meaningful entity embeddings during the initial training phase, making this approach impractical. We tackle this challenge by employing differentiable VQ~\cite{vqvae}. It enables dynamic adjustments of both entity embeddings and categorical code vectors from the quantization codebooks through recommendation tasks and quantization constraints, starting from their initial random states and resulting in a robust and stable form.

\input{figures/vq}

Moreover, it is crucial to select the appropriate level of detail for categories. Employing finely detailed categories could potentially result in data sparsity issues within the recommendation system, while adopting broader, coarse-grained categories may obscure significant differentiations among entities. In light of this challenge, we propose cascaded VQ to construct a category tree with varying levels of granularity. Unlike the single-layer category system commonly used in datasets, our cascaded approach creates a hierarchical taxonomy of categories, offering a more comprehensive representation of entities. 

% Second, choose the appropriate level of detail for categories is crucial. Fine-grained categories may result in sparse data for recommendations, while coarse-grained categories can obscure relevant entity distinctions. To mitigate this, we devise a cascaded clustering structure to capture varying granularities of categorical attributes.

To summarize, we introduce \model{}, an automatic category tree generation framework for ID-based recommendation, which offers several notable advantages and capabilities as outlined below.

% \begin{itemize}
\textbf{i. End-to-end framework.} Differing from the common multi-stage application of VQ in recommender systems~\cite{zhang2023query,rajput2023recommender}, we are the first to explore differentiable VQ as an end-to-end solution for category generation in a scenario without side information, i.e., ID-based recommendation.
    Such end-to-end training allows for refining and optimizing the categorization for both items and users to align with specific recommendation objectives.

\textbf{ii. Easy adoption and high adaptability.} \model{} is a pluggable module that can be conveniently integrated into both sequential and non-sequential recommendation models for accommodating different recommendation scenarios, including list completion, collaborative filtering, and click-through rate prediction.

\textbf{iii. Effectiveness.} We conduct a comprehensive evaluation of \model{} on multiple recommendation tasks, including list completion, collaborative filtering, and click-through rate prediction. The evaluation involves tem datasets and a comparison with 14 baseline methods. The results demonstrate the effectiveness of \model{}, showcasing significant improvements across most scenarios. For example, \model{} demonstrates a relative improvement of up to 21.41\% over state-of-the-art baselines in list completion tasks.

% \end{itemize}

%% file: figures/dove.tex
\begin{figure}[t]
    \centering
    \includegraphics[width=.9\linewidth]{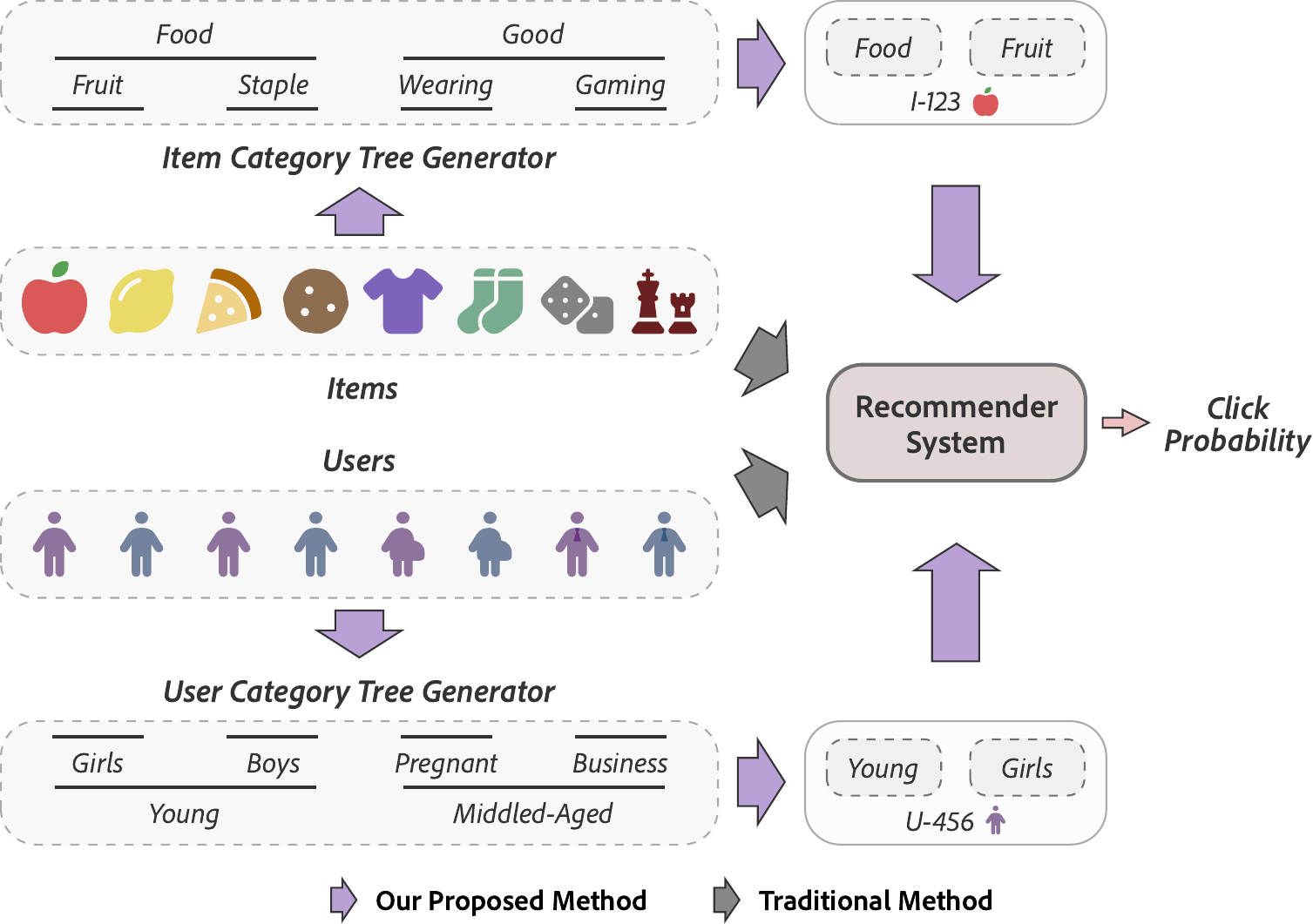}
    \caption{
    Illustration of our approach in learning category trees for ID-based recommendation. In contrast to traditional methods that solely offer item or user IDs to the recommender system, our approach involves implicit learning of user/item category trees. The category information, encoded as vectors, is subsequently integrated with the user/item ID and provided as input to the recommender system.
    % 输入给RS的不仅是ID，还有抽取到的category knowledge (vector implicit)
    %Integration of our proposed automatic category-tree generation module (i.e., \model{}) with the recommender system. Note that we represent entities concretely in the schematic for better visualization, but there is no actual side information provided. In practical situations, each node in the category tree is not human-readable, but is stored implicitly in the form of a vector.
    }\label{fig:illustration} 
\end{figure}

%% file: figures/vq.tex
% \begin{figure}[t]
%   \centering
%   \begin{subfigure}[b]{\linewidth}
%     \centering
%     \includegraphics[width=\linewidth]{images/VQ-Previous.pdf}
%     \caption{\label{fig:vq-previous}}
%   \end{subfigure}
%   \\
%   \begin{subfigure}[b]{\linewidth}
%     \centering
%     \includegraphics[width=\linewidth]{images/VQ-Ours.pdf}
%     \caption{\label{fig:vq-ours}}
%   \end{subfigure}
%   \caption{\label{fig:vq-compare} (a) Previous three-stage vector quantization based recommender system. (b) Our proposed end-to-end vector quantization based recommender system. White arrows indicate the propagation of gradients while black arrows denote stop gradient.}
% \end{figure}

\begin{figure}[t]
    \centering
    \includegraphics[width=.9\linewidth]{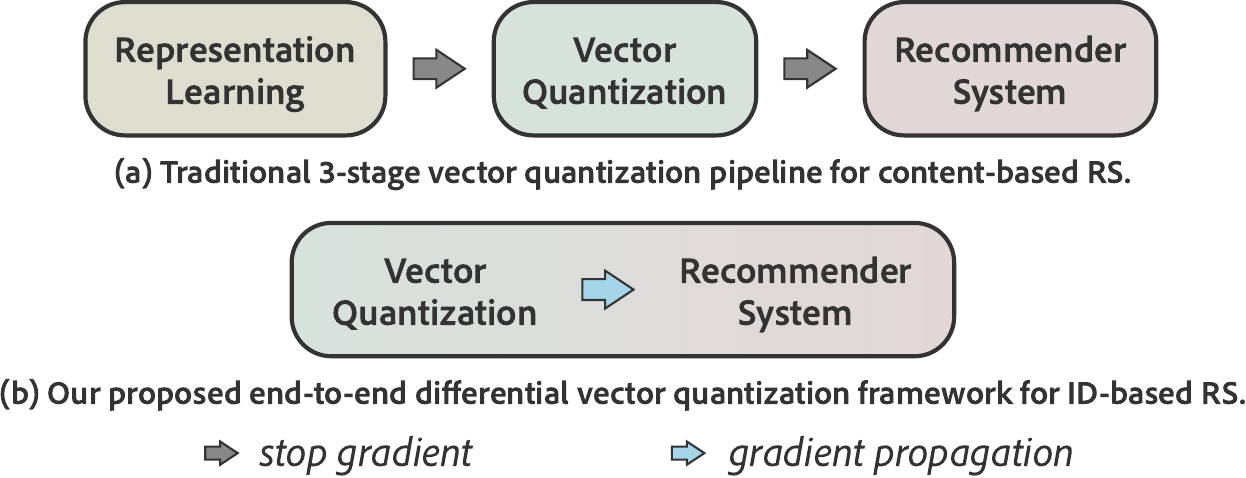}
    \caption{\label{fig:vq-compare} 
    %Paradigms for the vector quantization technique on the recommender system. 
    Comparison between (a) the traditional three-stage vector quantization pipeline for content-based recommendation and (b) our proposed end-to-end differential vector quantization framework for ID-based recommendation.
    }
\end{figure}

%% file: content/RelatedWork.tex
\section{Related Work}

\subsection{Vector Quantization}

Vector quantization (VQ) techniques~\cite{gray1984vector} map a large set of input vectors into a small set of vectors (i.e., a codebook), which have been widely studied in computer vision~\cite{xia2013joint,babenko2014additive,vqvae2} and speech coding~\cite{buzo1980speech,juang1982multiple} domains.
% In particular, VQ is extensively investigated to save bandwidth and storage when transmitting and storing images and speeches. For example, additive quantization (AQ)~\cite{babenko2014additive} compresses an image vector by representing it as a sum of several components from separate codebooks. The compressed vectors of AQ lead to an increased accuracy in image classification. In addition, VQ-VAE-2~\cite{vqvae2} uses the vector quantized variational autoencoder to generate large-scale high-resolution images. 
% \textbf{Vector Quantization for Recommendation.} \quad 
To date, only a few studies explore the potential of vector quantization in recommendation systems. 
% On one hand, vector quantization is applied to improve the efficiency of a recommender system (ICDM21 MASCOT, RecSys19 PQ-VAE). 
One line of research aims to improve recommendation efficiency~\cite{ko2021mascot,lian2020product,van2019pq}. The other line of research focuses on improving recommendation quality~\cite{rajput2023recommender,pan2021click}, as shown by the growing interest from researchers in recent years. The studies on enhancing recommendation quality can be categorized into two paradigms: the commonly used multi-stage approach~\cite{hou2023learning,zhang2023query, rajput2023recommender} and an end-to-end training~\cite{pan2021click} strategy.

To our knowledge, AQCL~\cite{pan2021click} is the only work using end-to-end training for quality improvement in ID-based recommendation, which leverages VQ to assist contrastive learning in the CTR prediction scenario. Differ from AQCL, our proposed \model{}, is the first to introduce VQ for learning categorical knowledge in ID-based recommendation.

\input{figures/cove}

\subsection{Recommender Systems}

Recommender systems have been extensively studied in various application scenarios including (1) list completion, which aims to continue the user-curated list by sequence generation, (2) collaborative filtering (CF) that makes recommendation based on user-item interactions, and (3) click-through rate (CTR) prediction, which is a crucial task in the ranking phase of the recommendation pipeline.

% \subsubsection{List Completion}
\textbf{List completion.} \quad 
Pioneer works based on Markov chain~\cite{mcfee2011natural,mcfee2012hypergraph,chen2012playlist} or neural networks~\cite{chen2018recsys, volkovs2018two,gatzioura2019hybrid,tran2019adversarial} are mostly proposed for automatic playlist continuation. 
% The pioneer works~\cite{mcfee2011natural,mcfee2012hypergraph,chen2012playlist} model item list via the Markov chain. 
% One line of research focuses on automatic playlist continuation~\cite{chen2018recsys, volkovs2018two,gatzioura2019hybrid,tran2019adversarial}. 
% However, these methods are often tailored for music playlists. 
In recent years, sequential recommenders~\cite{caser,gru4rec,bert4rec,car} have been proposed to generation items autoregressive for list completion task, while FANS~\cite{fans} uses non-autoregressive generation to improve both quality and efficiency.

\textbf{Collaborative filtering.} \quad 
% To overcome the scalability and sparsity issues in large-scale systems, matrix factorization~\cite{koren2009matrix} techniques and deep learning-based methods have been widely adopted to capture underlying preferences and characteristics for personalized recommendation.
Collaborative filtering (CF) is widely adopted in the matching phase of the recommendation pipeline. Traditional CF methods~\cite{itemcf,breese2013empirical,koren2009bellkor,lemire2005slope} employ neighborhood-based approaches and use similarity metrics to identify users or items with similar preferences, which face scalability and sparsity issues in large-scale systems. To overcome these limitations, matrix factorization~\cite{koren2009matrix} techniques have been widely adopted to capture underlying preferences and characteristics for personalized recommendation. More recently, deep learning-based methods~\cite{bpr,neumf,cfkg,lightgcn} have emerged to learn complex user-item interactions and capture nonlinear relationships.

\textbf{Click-through rate prediction.} \quad 
In recent years, deep learning-based CTR prediction models~\cite{cheng2016wide,dcn,deepfm,fibinet,mao2023finalmlp} have gained popularity. These models have demonstrated improved performance by leveraging the expressive power of neural networks to capture intricate patterns in user-item interactions.

%% file: figures/cove.tex
\begin{figure*}[ht]
    \centering
    \includegraphics[width=.8\linewidth]{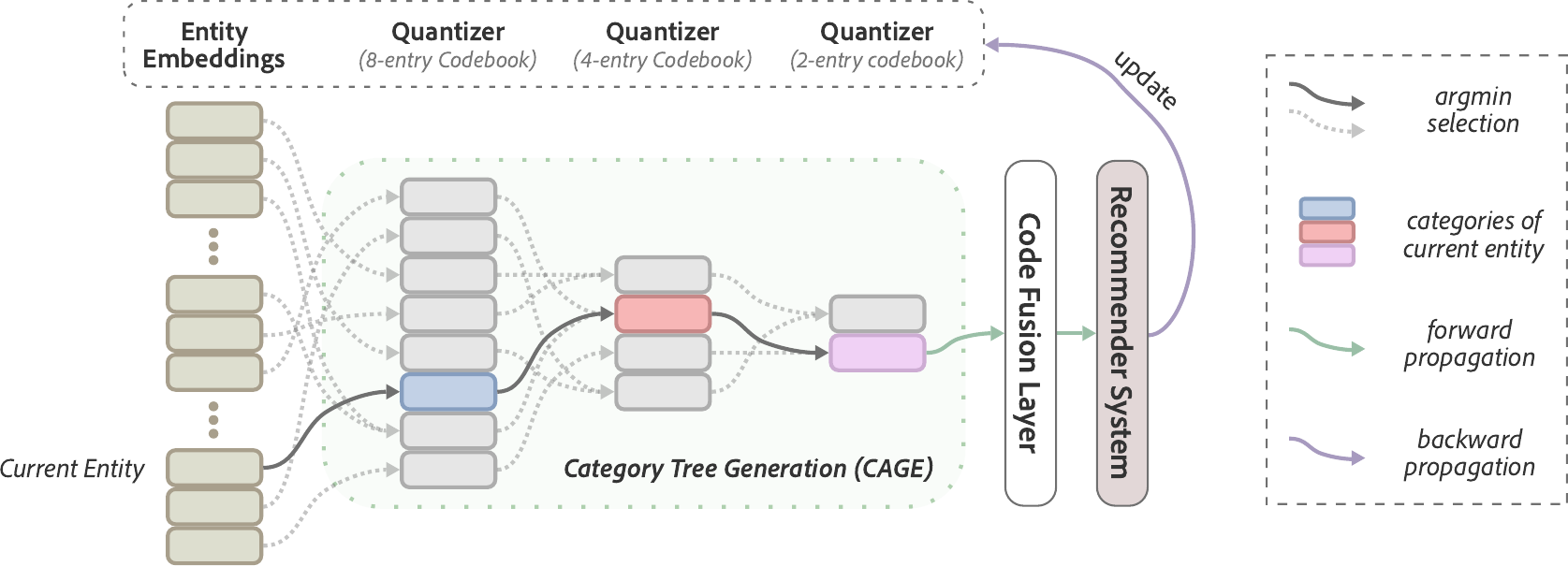}
    \caption{\label{fig:cove} 
    %An overview of the cascade code generator module.
    Overview of our proposed category tree generation framework (\model{}).}
\end{figure*}

%% file: content/Method.tex
\section{Proposed Framework: \model{}}

\textbf{Overview.} Figure~\ref{fig:cove} illustrates our automatic category tree generation (\model{}) framework, which is designed to enhance \emph{id-based} representations of both items and users.
It involves a series of cascaded vector quantizers for extracting category-aware information at multiple levels of granularity, thereby forming a tree-like architecture. Specifically, during training, each node (entity or category node) at the lower level searches for the closest category node at the higher level to establish a connection of belonging. Since the node embeddings are dynamically changed during the training process, the connections may be constantly updated. The searching process constructs a path from the bottom layer to the top layer for each entity node, which includes the multi-level category embeddings associated with the entity. Subsequently, the quantized multi-level code vectors are then fused and fed to the recommender system to facilitate downstream recommendation tasks. \model{} and the recommender system are trained together in an end-to-end manner. Once the training process is complete, all the connections are finalized, and the category tree is generated.

%During training, each node (entity or category node) at the lower level searches for the closest category node at the higher level to establish a connection of belonging. Note that since the node embeddings are dynamically changed during the training process, the connections may be constantly updated. The searching process constructs a path from the bottom layer to the top layer for each entity node, which includes the multi-level category embeddings associated with the entity. Subsequently, thes multi-level category embeddings are combined with the original entity embedding to form a category-aware entity embedding, which is then used as input to the recommendation module. We employ the Straight-Through Estimator (STE) technique to optimize the node embeddings during training. Once the training process is complete, all the connections are finalized, and the category tree is generated.

\subsection{Category Tree Construction}\label{sec:vq}

Before training, both the embedding vectors and code vectors in codebooks are randomly initialized. There are no connections between each entity and the nodes in the bottom-level codebook (the leftmost layer in~\autoref{fig:cove}), as well as between adjacent codebooks.

% Given an input vector $\mathbf{z}$, \model{} aims to extract cascaded code representations (i.e., blue, red, and yellow blocks depicted in Figure~\ref{fig:framework}).
% % It is implemented by a series of vector quantizers, a code fusion layer, and a differentiable back propagation mechanism.

\subsubsection{Searching with Vector Quantization}

Specifically, we establish links between adjacent layers using the vector quantization technique. Vector quantization~\cite{wu2019vector} targets at grouping similar vectors into clusters by representing them with a small set of prototype vectors. We use a vector quantizer to locate the code vector within a codebook that closely matches the input embedding. The code vector is anticipated to capture and represent the categorical information associated with the input embedding.
The vector quantizer includes a $k$-entry codebook $\mathbf{C} \in \mathbb{R}^{k \times d}$, where $k$ is the number of the code vectors and $d$ is the dimension of each code vector.

Given an input embedding $\mathbf{e} \in \mathbb{R}^d$, nearest neighbour search is performed to find the most similar code to $\mathbf{e}$ within $\mathbf{C}$:
\begin{equation}\label{eq:dist}
    j = \arg\min_{i \in \{1, 2, \ldots, k\}} \|\mathbf{e} - \mathbf{c}_i\|_2^2,
\end{equation}
where $\mathbf{c}_i (1\leq i\leq k)$ is any code vector in the codebook $\mathbf{C}$, and $j$ is the index of the matched code vector $\mathbf{c}_j$.
Note that the current matching pair $(\mathbf{e}, \mathbf{c}_j)$ is a \textbf{temporary result} that evolves as the training process continually adjusts entity embeddings and code vectors.

\subsubsection{Cascaded Linking Flow}

\model{} employs a series of cascaded vector quantizers to capture categorical information at multiple levels of granularity. Figure~\ref{fig:cove} shows an example with three quantizers. 

Let $H$ be the number of quantizers (or levels of granularity). Each quantizer ${Q}^{(i)}$ has a $v^i$-entry codebook $\mathbf{C}^{(i)}$, where $i = 1, 2, \ldots, H$. The quantizers are interconnected in a cascaded fashion, generating \emph{fine-to-coarse} code vectors, i.e., $v^i>v^j$ for $i<j$. Each quantizer ${Q}^{(i)}$ takes the output of the previous quantizer (i.e., ${Q}^{(i-1)}$) as input, creating a quantization flow defined as follows.
\begin{gather} \label{eq:vectorquantizer}
% \begin{aligned}
\mathbf{c}^{(i)} = {Q}^{(i)} \left(\mathbf{c}^{(i-1)}\right), \\
\mathbf{c}^{(0)} = \mathbf{e},
% \text{where}\quad \mathbf{c}^{(i-1)} &=
% 	\begin{cases}
% 		\mathbf{e}, & \mbox{if } i = 1 \\
% 		\mathbf{c}_\text{q}^{(i-1)}. & \mbox{else}
% 	\end{cases}
% \end{aligned}
\end{gather}
where $\mathbf{c}^{(i)}$ is the output of quantizer ${Q}^{(i)}$.
%\mathbf{E}^{(i)}_{y^{(i)}}$, 
%and $\mathbf{E}^{(i)}$ is the codebook of $i$-th layer vector quantizer.

% \input{figures/fuse}

\subsection{Code Fusion Layer}
After obtaining multi-level codes (or categories) $\mathbf{c}^{(i)} (i = 1, 2, \cdots, H)$, we employ an average pooling operation to combine them into a single vector:
\begin{gather}
\mathbf{\bar{c}} = \frac{1}{H} \sum^H_i \mathbf{c}^{(i)}.
\end{gather}
In addition, we use a weighted residual connection to add the original vector $\mathbf{e}$ to obtain the final category-aware representation $\mathbf{z}$:
\begin{gather}
\mathbf{z} = \mathbf{e} + \alpha \mathbf{\bar{c}}, \label{eq:weightedresnet}
\end{gather}
where $\alpha$ is a hyperparameter that balances the two terms. We use ``$f$'' to denote the aforementioned operations, i.e., $\mathbf{z} = f(\mathbf{e})$.
%hyperparameter that controls the contribution of the category-aware vector $\mathbf{\bar{z}}$, i.e., the weight of the residual connection.

\subsection{Tree Back Propagation}

Since the nearest neighbour search algorithm is not differentiable, we utilize the straight-through estimator (STE)~\cite{bengio2013estimating} to approximate the gradient of each quantizer.
Specifically, the gradient of the quantizer is approximated by the gradient of the identity function, which is defined as:
\begin{equation}
    \frac{\partial \mathbf{c}^{(i)}}{\partial \mathbf{c}^{(i-1)}} \approx \frac{\partial \mathbf{c}^{(i-1)}}{\partial \mathbf{c}^{(i-1)}} = \mathbf{I},
\end{equation}
where $\mathbf{I}$ is the identity matrix. Therefore, the quantization loss (encouraging the quantizer to select the closest vector in the codebook) can be defined as:
\begin{equation}
% \begin{aligned}
L_{\text{quant}} = \sum^H_i \left( \| sg [ \mathbf{c}^{(i-1)} ] - \mathbf{c}^{(i)} \|_2^2 \right),
% \end{aligned}
\end{equation}
where $sg$ is the stop gradient operation. Furthermore, we introduce a commitment loss that encourages the input embedding $\mathbf{c}^{(i-1)}$ to approach the currently matched code vector $\mathbf{c}^{(i)}$, which reduces the frequency of link changes, resulting in a smoother training process:
\begin{equation}
L_{\text{commit}} = \sum^H_i \left(\| \mathbf{c}^{(i-1)} - sg [ \mathbf{c}^{(i)} ] \|_2^2 \right).
\end{equation}

Finally, the overall tree generation loss can be defined by:
\begin{equation}
L_{\text{cage}} = L_{\text{quant}} + \beta L_{\text{commit}},
\end{equation}
where $\beta$ is a hyper-parameter that controls the trade-off between the two losses.

\subsection{End-to-end Training}

% As shown in Figure~\ref{fig:non-seq} and~\ref{fig:seq}, our \model{} can be easily integrated into a wide range of recommender systems, including non-sequential recommenders and sequential recommenders. More precisely, \model{} performs quantization on entity embeddings, extracting categorical information termed as cascaded code representation. Furthermore, it combines the initial entity embeddings, producing category-aware entity embeddings that seamlessly integrate within the recommender system. It's worth noting that we have designed both user \model{} and item \model{} modules, each meticulously tailored to process user and item embeddings, respectively. For the sequential recommendation context, user embedding will be learned by item history list, so only item \model{} is required.

As mentioned in Section~\ref{sec:vq}, the cascaded code vectors and entity embeddings are both initialized randomly prior to training. Initially, entity embeddings lack meaningful information, leading to insignificant quantization outcomes. As training progresses, the \model{} module and the recommender model are jointly optimized through an external recommendation task (i.e., recommendation loss), gradually imbuing entity embeddings with semantic context. Furthermore, internal tree generation loss, $L_\text{cage}$, is introduced to enhance the clustering effectiveness of the codebook. The enriched category information (code representation) subsequently contributes to improved recommendation performance for entity embeddings in subsequent training batches. This cyclic iteration results in a double-helix refinement process, where the codebook and entity embeddings continuously enhance their representation learning throughout the training process. 
% Hence, we refer to such training scheme as ``co-evolving training'' and name our model ``\textbf{COVE}'' (\textbf{co}-evolving \textbf{ve}ctor quantization).
% Specifically, the loss objective of the sequential and non-sequential recommenders can be respectively defined as:
% \begin{gather}\label{eq:general}
% L_\text{seq} = L_{\text{rec}} + \omega_{q} L_\text{quant}^\text{item}, \\
% L_\text{non\_seq} = L_\text{seq} + \omega_{q} L_\text{quant}^\text{user},
% \end{gather}
% where the hyperparameter $\omega_{q}$ balances the objectives of accurate recommendation and codebook quantization during the training process.

%the \model{} module is placed before the recommender, which takes the item vector $\mathbf{z}$ as input and outputs the category-aware item vector $\mathbf{z}_\text{c}$, which is then fed into the recommender.

% In Algorithm~\ref{algo:integration}, we provide a clear demonstration of how \model{} can be easily integrated into an existing recommender system with just three lines of code modification.

%have encapsulated the \model{} module, which can be seamlessly integrated into any recommenders within three lines of code modification~\footnote{The code is available at \url{https://anonymous.4open.science/r/CIKM23-C5/}}.

%% file: content/Application.tex
\section{Applications}

\model{} can be effortlessly and seamlessly integrated into a variety of recommendation models to enhance recommendation performance, making it highly adaptable and suitable for a wide range of recommendation scenarios.  In the following, we demonstrate how \model{} can be applied to non-sequential (e.g., collaborative filtering and CTR prediction) and sequential recommenders (e.g., list completion).

\subsection{\model{} for Non-sequential Recommenders: Incorporating both Item and User \model{}}

\subsubsection{Scenario 1: Collaborative Filtering.}

Given a set of users $\mathcal{U}$ and a set of items $\mathcal{V}$, the collaborative filtering agent~\cite{lightgcn,cfkg,bpr,neumf} aims to estimate the user-item interaction matrix $\mathbf{R}$, where each weight $r_{ui}$ represents the preference or rating of user $u$ for item $i$. The matrix $\mathbf{R}$ is typically sparse, as not all users rate or interact with all items. Therefore, the task is to fill in these missing weights by existing ratings.

\subsubsection{Scenario 2: Click-Through Rate Prediction.}

In contrast to collaborative filtering, which is typically employed in the matching phase of the recommender pipeline, CTR prediction is a ranking-based task that aims to predict the probability of a user clicking on a particular item.
The input to the CTR model is a user-item pair, and the output is a probability score indicating the likelihood of the user clicking on the item. Existing deep CTR prediction models~\cite{dcn,deepfm,pnn,fibinet} are typically designed to learn feature interactions from the raw input features such as user ID, item ID, and other statistical features if exists.

\subsubsection{Integration.}

In the non-sequential scenario, both user and item representations are obtained as embedding matrices and play integral roles in the training process. Therefore, we can incorporate two \model{} modules (i.e., Item \model{} $f^{(i)}$ and User \model{} $f^{(u)}$) on dual sides to extract hierarchical category knowledge, denoted as:
\begin{gather}
    \mathbf{z}^{(u)} = f^{(u)} \left(\mathbf{e}^{(u)}\right), 
    \mathbf{z}^{(i)} = f^{(i)} \left(\mathbf{e}^{(i)}\right).
\end{gather}
Finally, the loss function can be calculated by:
\begin{gather}
    L_\text{rec} = \Phi \left(\mathbf{z}^{(u)}, \mathbf{z}^{(i)}, l\right), \\
    L = L_\text{rec} + \omega_{q}\left(L_\text{cage}^{(u)} + L_\text{cage}^{(i)}\right), ~\label{eq:wq}
\end{gather}
where $\Phi$ is the non-sequential recommender, $l$ is the label, and $\omega_{q}$ is a hyperparameter balancing the internal tree generation loss and external recommendation loss.

\subsection{\model{} for Sequential Recommenders: Incorporating Item \model{} with an Additional Category Tree Classification Loss}

\subsubsection{Scenario: List Completion}

Given a set of item vocabulary $\mathcal{V} (v^0 = |\mathcal{V}|)$ and a user curated list $\mathbf{x} = [ x_1, x_2, \cdots, x_{|\mathbf{x}|}]$ ($x_i\in \mathcal{V}$), the list completion agent~\cite{car,caser,sasrec,fans} is required to predict an item sequence $\mathbf{y} = [ y_1, y_2, \cdots, y_{|\mathbf{y}|}]$ ($y_i\in \mathcal{V}$) that is a subsequent of $\mathbf{x}$, which can be formulated as maximizing the probability
\begin{equation}
    p\left(\mathbf{y}^\prime = \mathbf{y}|\mathbf{x} \right),
\end{equation}
where $\mathbf{y}^\prime$ represents any possible list of length $|\mathbf{y}|$.

Different from non-sequential recommenders which generate a scalar score, the output of the list completer is a prediction item vector $\mathbf{\bar{z}}$. More precisely, the item completer undergoes training using the item prediction task\footnote{This task can take the form of predicting the next item for autoregressive methods~\cite{caser,car} or a masked item prediction task~\cite{fans} for non-autoregressive methods.}. Therefore, a classification module is designed to infer the probability distribution over the item vocabulary by the softmax function for each prediction item vector:
\begin{equation}
    g^{0}: \mathbb{R}^d \rightarrow \mathbb{R}^{v^{0}}.
\end{equation}

% In previous list completion approaches, item prediction tasks such as next item prediction and masked item prediction are employed as training objectives. Our \model{} module is compatible with both of these tasks.

% \subsubsection{Original Flow}

% The general list completion model takes an item sequence $\mathbf{s} = [s_1, s_2, \cdots, s_l]$ as input, where $\mathbf{s}$ can be a subsequence of $\mathbf{x}$ and $l$ represents the sequence length.
% After being passed through an embedding layer, the discrete item sequence is transformed into the dense sequence embeddings $\mathbf{Z} \in \mathbb{R}^{l \times d}$.
% Subsequently, the sequential encoder will capture the contextual information in the sequence embeddings and generate a prediction sequence $\mathbf{Z}_\text{p} \in \mathbb{R}^{l \times d}$.
% By applying item prediction tasks such as next item prediction~\cite{car,caser,sasrec} and masked item prediction~\cite{bert4rec,fans}, a classification module ${C^{0}}: \mathbb{R}^d \rightarrow \mathbb{R}^{v^{0}}$ is used to infer the probability distribution over the item vocabulary by the softmax function for each prediction item vector $\mathbf{z}_\text{p} \in \mathbf{Z}_\text{p}$.

% \input{figures/completion}

\subsubsection{Integration.}

In the sequential scenario, since the user representation is derived by fusing the historical item list, we only need to insert a single \model{} module for item categorization.
Additionally, the category tree generated by our \model{} naturally serves as a valuable aid for the item prediction task.

Assuming that the ground truth label for the item vector to be predicted, $\bar{\mathbf{z}}$, is the $y^{(0)}$-th item, and its current embedding is $\mathbf{z}$. We can start by using Item \model{} to obtain pseudo-labels (i.e., \textbf{\textit{code indices}} in the multi-level codebooks, denoted as $y^{(i)}, i = 1, 2, \ldots, H$) for the current embedding $\mathbf{z}$. Subsequently, we design an auxiliary \textbf{\textit{tree classification task}} that encourages the current prediction vector $\bar{\mathbf{z}}$ to predict the category it corresponds to in each layer of the category tree which multiple node classification module:
\begin{equation}
    g^{i}: \mathbb{R}^d \rightarrow \mathbb{R}^{v^{i}}, i = 1, 2, \ldots, H.
\end{equation}
Such auxiliary task further strengthens the connection between items and categories, leading to more precise predictions. Then, we proceed with the multi-level classification training and the loss function can be defined as:
\begin{gather}
    L_\text{item} = g^{0}\left(\mathbf{z}\right)_{y^{(0)}}, \\
    L_\text{tree} = \frac{1}{H} \sum^H_i g^{i}\left(\mathbf{z}\right)_{y^{(i)}}.
\end{gather}

Finally, the overall recommendation loss function is:
\begin{gather}
    L_\text{rec} = L_\text{item} + \omega_\text{c} L_\text{tree}, \\
    L = L_\text{rec} + \omega_{q} L_\text{cage},
\end{gather}
where $\omega_\text{c}$ is a hyperparameter that controls the importance of the tree classification loss, and please refer to ~\autoref{eq:wq} for $\omega_\text{q}$.

%% file: content/Experiment.tex
\section{Experiment}

\subsection{Experimental Setup}

\subsubsection{Datasets.}
We conducted offline experiments on three recommendation tasks, namely list completion, collaborative filtering (CF), and click-through rate (CTR) prediction. For the list completion task, we use three real-world datasets: Zhihu, Spotify, and Goodreads, which were crawled and compiled by~\cite{car}.
% Spotify\footnote{\url{https://www.spotify.com}} is a music streaming service. Zhihu\footnote{\url{https://www.zhihu.com/}} is a Chinese question-and-answer website. Goodreads\footnote{\url{https://www.goodreads.com/}} is a social cataloging website dedicated to searching and sharing books. 
For the collaborative filtering task, we utilize five public datasets: Amazon Toys, Amazon Kindle Store, Amazon Phones, Amazon Grocery, and MovieLens~\cite{movielens} (1M version). 
Regarding the CTR prediction task, 
we employ two public datasets: MIND~\cite{mind} (small version) and MovieLens~\cite{movielens} (100K version). 
% MIND\footnote{\url{https://www.microsoft.com/en-us/research/project/mind/}} is a large-scale news recommendation dataset. MovieLens\footnote{\url{https://grouplens.org/datasets/movielens/}} is a well-known movie recommendation dataset. 
Please refer to Appendix~\ref{sec:statistics} for the dataset statistics.

\subsubsection{Preprocessing.}
For the list completion task, we adopt the data preprocessing steps proposed by~\cite{fans}. 
We iteratively perform the following two operations until the data no longer changes: 1) remove items with a frequency less than 10 from all lists; 2) truncate or filter the item list according to the maximum and minimum lengths specific to each dataset. Furthermore, we uniformly divide a qualifying list into two segments, namely the input and target lists. The lists are then partitioned into training, validation, and testing sets using an 8:1:1 ratio.

For the CF and CTR prediction datasets, we only use the user-item interaction data without any additional information.
To be specific, for the MIND dataset, user historical behaviors are transformed into a list of user-item pairs, which are subsequently included in the training set. 
More details about the dataset preprocessing will be provided in the public code repository upon accepted.

\subsubsection{Baselines and Variants of Our Method.}

\textbf{List completion.} We take the state-of-the-art sequential recommendation methods and the item list completion models as baselines, including Caser~\cite{caser}, GRU4Rec~\cite{gru4rec}, SASRec~\cite{sasrec}, BERT4Rec~\cite{bert4rec}, CAR~\cite{car} and FANS~\cite{fans}. We integrate \model{} into BERT4Rec and FANS to obtain \model{}$_\text{BERT4Rec}$ and \model{}$_\text{FANS}$ models, respectively.

% \model{}$_\text{BERT4Rec}$ and \model{}$_\text{FANS}$ are the integration of \model{} into BERT4Rec and FANS, respectively.

It is worth noting that FANS~\cite{fans} pre-extracts categorical item features based on the curated item lists among training, validation, and testing sets. These categorical knowledge is also added into baseline models for a fair comparison in the FANS paper.
Since we learn the cascaded categorical features in an end-to-end manner, we do not use the pre-extracted categorical entity features in our experiments for both our method variants and baselines.

\textbf{Collaborative filtering.} We compare our method with representative CF models as baselines, including BPRMF~\cite{bpr}, NeuMF~\cite{neumf}, CFKG~\cite{cfkg} and LGCN~\cite{lightgcn}
We integrate our proposed \model{} module into these baselines and denote them as \model{}$_\text{BPRMF}$, \model{}$_\text{NeuMF}$, \model{}$_\text{CFKG}$, and \model{}$_\text{LGCN}$, respectively.

\input{tables/completion}
\input{tables/cf}

\textbf{Click-through rate prediction.} We compare our method with the widely used and state-of-the-art deep CTR models, including DeepFM~\cite{deepfm}, DCN~\cite{dcn}, FiBiNET~\cite{fibinet}, and FinalMLP~\cite{mao2023finalmlp}.
We integrate our proposed \model{} module into these baselines and denote the integrated models as \model{}$_\text{DeepFM}$, \model{}$_\text{DCN}$, \model{}$_\text{FinalMLP}$, and \model{}$_\text{FiBiNET}$, respectively.

\subsubsection{Evaluation Protocols.}
We follow the common practice~\cite{shi2019deep} to evaluate the effectiveness of recommendation models with the widely used metrics, i.e., Normalized Discounted Cumulative Gain~\cite{ndcg} (NDCG@k) and Hit Ratio (HR@k). In this work, we set $k = \{5, 10\}$.

% \subsubsection{Implementation Details.}

% Due to the space limit, please refer to the Appendix section.
% During training, we adopt the Adam optimizer as the gradient descent algorithm.
% For all models, the embedding dimension is set to 64. \textbf{For the list completion task}, we set the batch size to 256 and the learning rate to 0.01 following~\cite{fans}. We use 3 Transformer layers for all Transformer-based models and 3 hidden layers for the GRU4Rec model. For the Caser model, we follow the original implementation and settings, and set the max sequence length to 5. We set the number of attention heads to 8 for all Transformer-based methods on the three datasets of list completion. \textbf{For the collaborative filtering task}, we set the batch size to 1024 and the learning rate to 0.001. For the LGCN model, we set the number of GCN layers to 3. \textbf{For the CTR prediction task}, we set the batch size to 5000, the learning rate to 0.001, the number of DNN layers to 3, the size of each hidden layer to 1000, and the dropout rate to 0.1 for all models.
% For the DCN model, we set the number of cross layers to 3.
% For the FiBiNET model, we set the number of feature interaction blocks to 3.

% We carefully tune the hyper-parameters of all models on the validation set and report the best results achieved on the test set. The results are averaged over 5 runs. Due to space constraints, we will furnish the details in future publications.
% All the methods were trained using NVIDIA GeForce RTX 3090 with 24GB memory.

\input{tables/ctr}
\input{tables/layers}

\subsection{Main Results}

\textbf{List Completion.} Table~\ref{tab:completion} presents a comparison of the state-of-the-art sequential recommenders with our proposed \model{} variants on the list completion task.
Based on the results, we can make the following observations.
\textbf{Firstly}, for both autoregressive and non-autoregressive models, our proposed \model{} module can significantly improve the performance of the baseline models.
For example, \model{}$_\text{BERT4Rec}$ can achieve an average improvement of 38\% and 31\% in terms of NDCG@5 and HR@5 among all datasets, compared with BERT4Rec.
\textbf{Secondly}, since the FANS models leverage item category information in their design, they outperform other autoregressive baselines.
However, our \model{}-integrated variant \model{}$_\text{FANS}$ can still achieve better performance than FANS, which implies that the end-to-end training models utilizing differentiable vector quantization can effectively learn improved clustering features compared to the word2vec+kmeans~\cite{fans} approach that relies on pre-extracted features.
\textbf{Thirdly}, in the Spotify dataset, the performance of the CNN-based Caser model is better than Transformer-based BERT4Rec model, which is aligned with the observation in~\cite{fans}.
One possible reason is that the local knowledge of the Spotify dataset is more important than the global information.

\textbf{Collaborative Filtering.} Table~\ref{tab:cf} displays the results of the popular CF models, along with our proposed \model{} variants on the collaborative filtering task. From the results, we can observe that our proposed \model{} consistently enhances the performance on the five datasets, resulting in significant improvements compared to the baseline models.
%For both general and graph-based CF models, our proposed \model{} module can significantly improve the performance of the baseline models on the two datasets.

\textbf{Click-Through Rate Prediction.} Table~\ref{tab:ctr} shows the results of the widely-used CTR prediction models and our proposed \model{} variants on the CTR prediction task.
Based on the results, we can find that among all CTR prediction models, our \model{} variants outperform the baseline models.

\input{plots/dual/main}
\input{plots/hyper}

\subsection{Ablation Study}~\label{subsec:ablation}

\textbf{Structure of the Category Tree.} We study the effects of the number of layers and the number of entries (i.e., codebook size) in \model{}.
We vary the number of layers from 1 to 3 and the number of entries within a range from 10 to 8,000.
We fix other hyper-para\-meters and report the results of \model{}$_\text{BERT4Rec}$ and \model{}$_\text{FANS}$.
As illustrated in Table~\ref{tab:layers}, we conduct experiments on the Zhihu and Goodreads datasets.
From the results, we can make the following observations.
\textbf{Firstly}, the best results of two-layer \model{} variants are better than those of one-layer \model{} variants on both datasets, indicating that \model{} can effectively capture the hierarchical category information to further improve the entity representations.
\textbf{Secondly},  different variants prefer different numbers of entries.
For example, on the Zhihu dataset, \model{}$_\text{BERT4Rec}$ prefers a small number of entries in the first layer (i.e., 200), while \model{}$_\text{FANS}$ prefers a large number of entries in the same layer (i.e., 500).
\textbf{Thirdly}, different datasets prefer different numbers of entries.
For example, for the \model{}$_\text{BERT4Rec}$ variant, the best number of entries is 20 on the Zhihu dataset and 50 on the Goodreads dataset.
\textbf{Fourthly}, as the number of entries increases, the performance of \model{} variants first increases and then decreases.
One possible reason is that a small number of entries may exhibit boundary effects, and as the entry size increases, the boundaries of the clusters gradually become blurred.
However, when the number of entries is too large, the number of entities in each entry is too small, which may lead to insufficient learning of categorical feature.
Moreover, the layer and entry numbers need to be carefully adjusted, otherwise it may lead to negative effects.

\textbf{Effectiveness of Item/User \model{}.} We also test the effectiveness of the dual \model{} (i.e., using both user and item \model{}) in both CF and CTR prediction scenario. As shown in Figure~\ref{fig:dual}, the results prove that both user and item \model{} could boost the performance of baselines.

\subsection{Impact of Hyper-parameters}

We explore the impacts of the residual connection weight $\alpha$, the quantization commitment cost $\beta$, the quantization loss weight $\omega_\text{q}$, and the codebook classification loss weight $\omega_\text{c}$. The experiments are conducted on two list completion datasets, i.e., Zhihu and Goodreads. Based on the results from Section~\ref{subsec:ablation}, we take the best \model{} configuration of the \model{}$_\text{FANS}$ model, i.e., (500, 10) for the Zhihu dataset and (500, 50) for the Goodreads dataset. Based on the results from Figure~\ref{fig:hyper}, we can make the following observations.
\textbf{Firstly}, the performance of baselines (i.e., when hyper-parameters are set to $0$) is inferior to the most of the cases, indicating the effectiveness of these hyper-parameters. \textbf{Secondly}, different datasets achieve the best performance at different hyper-parameter settings. For example, the Zhihu dataset reaches the best performance at $\alpha=0.6$, while for the Goodreads dataset, $\alpha=1.0$.
\textbf{Thirdly}, unlike the computer vision domain where the quantization commitment cost $\beta$ is usually set to $0.25$~\cite{vqvae}, in the recommendation domain, a higher $\beta$ (i.e., $1.0$ for the Zhihu dataset or $0.50$ for the Goodreads dataset) gets a higher performance.
\textbf{Fourthly}, due to the equivalent performance shown by the hyperparameters when set to 1 in the Figure~\ref{fig:hyper} (a-d) (e.g., the performance of $\alpha=1$ in  Figure~\ref{fig:hyper} (a) is equivalent to that of $\beta=1$ in Figure~\ref{fig:hyper} (b)), we can assess the performance when the hyperparameters are set to 0 by examining the range on the vertical axis (e.g., comparing the performance of $\alpha=0$ in Figure~\ref{fig:hyper} (a) with that of $\beta=0$ in Figure~\ref{fig:hyper} (b)). A wider range signifies a larger disparity between the performance at 0 and 1 for the hyperparameters. This indicates that when this particular hyper-parameter is set to 0, the resulting effect is poorer, highlighting its greater significance. Therefore, we can observe that the ranking of importance for these four hyperparameters is: $\omega_\text{q} > \beta > \alpha \approx \omega_\text{c}$.
Similarly, according to Figure~\ref{fig:hyper} (e-h) for the Goodreads dataset, the importance ranking is: $\omega_\text{q} \approx \beta > \alpha > \omega_\text{c}$.

% \input{figures/visual}

% \subsection{Visualization}

% Here, we demonstrate the clustering quality of our \model{}. We take the MIND dataset which has ground truth category labels in the experiments. We set the levels of vector quantizer to 1 (i.e., $H = 1)$ and the codebook size to 20 (i.e., $v^1 = 20$).

% Before training (we use the DeepFM model as the backbone), we randomly select one real category. After each training epoch, we calculate the relative proportion of news articles that each codebook entry contains for the current category. This operation will yield an array of 20 numbers, and their sum equals one.
% After training, we aggregate the arrays collected at the end of each epoch into a two-dimensional array and create a HeatMap as shown in~\autoref{fig:ex-visual}. We can observe that news articles for the current category are dispersed among different clusters. As training progresses, these news articles quickly converge into two clusters and stabilize in the subsequent phases, which demonstrates the effectiveness of the categorization of our \model{}.

% \input{plots/alpha/alpha}
% \input{plots/beta/beta}
% \input{plots/omega_code/omega}

%% file: tables/completion.tex
\begin{table*}[]

\centering
\renewcommand\arraystretch{1.15}

\resizebox{.875\linewidth}{!}{

\begin{tabular}{cc|cccccc|cccc}
\toprule
\multicolumn{2}{c|}{\textbf{Models}} & \makecell{\textbf{Caser} \\ \small{(\citeyear{caser})}}
& \makecell{\textbf{GRU4Rec} \\ \small{(\citeyear{gru4rec})}}
& \makecell{\textbf{SASRec} \\ \small{(\citeyear{sasrec})}}
& \makecell{\textbf{CAR} \\ \small{(\citeyear{car})}}
& \makecell{\textbf{BERT4Rec} \\ \small{(\citeyear{bert4rec})}}
& \makecell{\textbf{\model{}}$_\text{BERT4Rec}$ \\ \small{(ours)}}
& \makecell{\textbf{FANS*} \\ \small{(\citeyear{fans})}}
& \makecell{\textbf{FANS}$_\text{TSC}*$ \\ \small{(\citeyear{fans})}}
& \makecell{\textbf{\model{}}$_\text{FANS}$ \\ \small{(ours)}}
& \textbf{Imp.} \\ 
\midrule
\multirow{4}{*}{
\rotatebox{90}{\textbf{Zhihu}}
}
& N@5 & 0.0066 & 0.0058 & 0.0070 & 0.0138 & 0.0194 & 0.0220 & 0.0256 & 0.0232 & 0.0301 & 17.58\% \\
 & N@10 & 0.0107 & 0.0085 & 0.0111 & 0.0211 & 0.0290 & 0.0305 & 0.0389 & 0.0337 & 0.0428 & 10.03\% \\
 & HR@5 & 0.0958 & 0.0819 & 0.1043 & 0.1739 & 0.2140 & 0.2333 & 0.2857 & 0.2670 & 0.3034 & 6.20\% \\
 & HR@10 & 0.0958 & 0.1597 & 0.2020 & 0.3101 & 0.3758 & 0.3987 & 0.4819 & 0.4604 & 0.4859 & 0.83\% \\
\midrule
\multirow{4}{*}{
\rotatebox{90}{\textbf{Spotify}}
}
& N@5 & 0.0133 & 0.0041 & 0.0147 & 0.0166 & 0.0178 & 0.0202 & 0.0313 & 0.0315 & 0.0352 & 11.75\% \\
 & N@10 & 0.0183 & 0.0057 & 0.0203 & 0.0231 & 0.0257 & 0.0298 & 0.0461 & 0.0438 & 0.0519 & 12.58\% \\
 & HR@5 & 0.1924 & 0.0805 & 0.2062 & 0.2384 & 0.2638 & 0.3242 & 0.4071 & 0.3992 & 0.4385 & 7.71\% \\
 & HR@10 & 0.2764 & 0.1236 & 0.3159 & 0.3630 & 0.3991 & 0.4559 & 0.5927 & 0.5552 & 0.6282 & 5.99\% \\
\midrule
\multirow{4}{*}{
\rotatebox{90}{\textbf{Goodreads}}
}
& N@5 & 0.0089 & 0.0053 & 0.0061 & 0.0129 & 0.0108 & 0.0130 & 0.0334 & 0.0293 & 0.0399 & 19.46\% \\
 & N@10 & 0.0123 & 0.0068 & 0.0084 & 0.0188 & 0.0160 & 0.0180 & 0.0467 & 0.0418 & 0.0567 & 21.41\% \\
 & HR@5 & 0.1064 & 0.0856 & 0.1005 & 0.1712 & 0.1634 & 0.1829 & 0.3819 & 0.3268 & 0.4275 & 11.94\% \\
 & HR@10 & 0.1492 & 0.1252 & 0.1582 & 0.2672 & 0.2678 & 0.2678 & 0.5149 & 0.4514 & 0.5473 & 6.29\% \\
\bottomrule
\end{tabular}

}

\caption{Effectiveness of \model{} in list completion. We bold the best results. Asterisk symbol * indicates that the method uses pre-extracted categorical features which are learnt from the overall dataset including the test set.}
\label{tab:completion}

\end{table*}

%% file: tables/cf.tex
\begin{table*}[]
\centering
\renewcommand\arraystretch{1.15}

\resizebox{.875\linewidth}{!}{
\begin{tabular}{cc|ccc|ccc|ccc|ccc}
\toprule
\multicolumn{2}{c|}{\textbf{Models}}
& \makecell{\textbf{BPRMF} \\ \small{(\citeyear{bpr})}}
& \makecell{\textbf{\model{}}$_\text{BPRMF}$ \\ \small{(ours)}}
& \textbf{Imp.}
& \makecell{\textbf{NeuMF} \\ \small{(\citeyear{neumf})}}
& \makecell{\textbf{\model{}}$_\text{NeuMF}$ \\ \small{(ours)}}
& \textbf{Imp.}
& \makecell{\textbf{CFKG} \\ \small{(\citeyear{cfkg})}}
& \makecell{\textbf{\model{}}$_\text{CFKG}$ \\ \small{(ours)}}
& \textbf{Imp.}
& \makecell{\textbf{LGCN} \\ \small{(\citeyear{lightgcn})}}
& \makecell{\textbf{\model{}}$_\text{LGCN}$ \\ \small{(ours)}}
& \textbf{Imp.}
 \\ 
\midrule

\multirow{4}{*}{
\rotatebox{90}{\textbf{Toys}}
}
 & \textbf{N@5}
    & 0.2284 & \textbf{0.2352} & 2.98\%  
    & 0.1829 & \textbf{0.1903} & 4.05\% 
    & 0.2713 & \textbf{0.2824} & 4.09\% 
    & 0.2338 & \textbf{0.2401} & 2.69\% \\
 & \textbf{N@10} 
    & 0.2595 & \textbf{0.2702} & 4.12\%  
    & 0.2160 & \textbf{0.2218} & 2.69\%  
    & 0.3075 & \textbf{0.3162} & 2.83\% 
    & 0.2696 & \textbf{0.2755} & 2.19\% \\
 & \textbf{HR@5}    
    & 0.3131 & \textbf{0.3292} & 5.14\%  
    & 0.2589 & \textbf{0.2681} & 3.55\%  
    & 0.3703 & \textbf{0.3787} & 2.27\% 
    & 0.3216 & \textbf{0.3333} & 3.54\% \\
 & \textbf{HR@10}   
    & 0.4095 & \textbf{0.4377} & 6.89\%  
    & 0.3616 & \textbf{0.3661}  & 1.24\%  
    & 0.4826 & \textbf{0.4829} & -
    & 0.4328 & \textbf{0.4428}  & 2.31\% \\
    
\midrule
\multirow{4}{*}{
\rotatebox{90}{\textbf{Kindles}}
}
 & \textbf{N@5}
    & 0.4855 & \textbf{0.5112} & 5.29\% 
    & 0.4341 & \textbf{0.4431} & 2.07\%  
    & 0.4449 & \textbf{0.4686} & 5.33\% 
    & 0.5105 & \textbf{0.5123} & 0.35\% \\
 & \textbf{N@10} 
    & 0.5182 & \textbf{0.5443} & 5.04\%  
    & 0.4708 & \textbf{0.4789} & 1.72\%  
    & 0.4842 & \textbf{0.5062} & 4.54\% 
    & 0.5446 & \textbf{0.5476} & 0.55\% \\
 & \textbf{HR@5}
    & 0.6159 & \textbf{0.6494} & 5.44\%  
    & 0.5646 & \textbf{0.5711} & 1.15\%  
    & 0.5871 & \textbf{0.6070} & 3.39\% 
    & 0.6474 & \textbf{0.6573} & 1.53\% \\
 & \textbf{HR@10}   
    & 0.7168 & \textbf{0.7511} & 4.79\%  
    & 0.6778 & \textbf{0.6816} & 0.56\%  
    & 0.7084 & \textbf{0.7232} & 2.09\% 
    & 0.7526 & \textbf{0.7660}  & 1.78\% \\
\midrule
\multirow{4}{*}{\rotatebox{90}{\textbf{Phones}}} & N@5 & 0.2595 & \textbf{0.2693} & 3.78\% & 0.2145 & \textbf{0.2213} & 3.17\% & 0.2756 & \textbf{0.2840} & 3.05\% & 0.2849 & \textbf{0.2929} & 2.81\% \\
 &  N@10 & 0.2938 & \textbf{0.3043} & 3.57\% & 0.2479 & \textbf{0.2544} & 2.62\% & 0.3125 & \textbf{0.3205} & 2.56\% & 0.324 & \textbf{0.3310} & 2.16\% \\
 &  HR@5 & 0.3536 & \textbf{0.3687} & 4.27\% & 0.2989 & \textbf{0.3058} & 2.31\% & 0.3773 & \textbf{0.3875} & 2.70\% & 0.3933 & \textbf{0.4025} & 2.34\% \\
 &  HR@10 & 0.4598 & \textbf{0.4770} & 3.74\% & 0.4025 & \textbf{0.4085} & 1.49\% & 0.4918 & \textbf{0.5004} & 1.75\% & 0.5146 & \textbf{0.5205} & 1.15\% \\
\midrule
\multirow{4}{*}{\rotatebox{90}{\textbf{Grocery}}} & N@5 & 0.2435 & \textbf{0.2488} & 2.18\% & 0.2210 & \textbf{0.2285} & 3.39\% & 0.2822 & \textbf{0.2907} & 3.01\% & 0.2619 & \textbf{0.2707} & 3.36\% \\
 &  N@10 & 0.2774 & \textbf{0.2852} & 2.81\% & 0.2576 & \textbf{0.2659} & 3.22\% & 0.3247 & \textbf{0.3330} & 2.56\% & 0.3006 & \textbf{0.3108} & 3.39\% \\
 &  HR@5 & 0.3472 & \textbf{0.3572} & 2.88\% & 0.3265 & \textbf{0.3303} & 1.16\% & 0.3981 & \textbf{0.4115} & 3.37\% & 0.3784 & \textbf{0.3907} & 3.25\% \\
 &  HR@10 & 0.4517 & \textbf{0.4689} & 3.81\% & 0.4389 & \textbf{0.4450} & 1.39\% & 0.5292 & \textbf{0.5420} & 2.42\% & 0.4972 & \textbf{0.5140} & 3.38\% \\
\midrule
\multirow{4}{*}{\rotatebox{90}{\textbf{MovieLens}}} & N@5 & 0.3511 & \textbf{0.3553} & 1.20\% & 0.3425 & \textbf{0.3479} & 1.58\% & \textbf{0.3674} & 0.3665 & - & 0.1284 & \textbf{0.1350} & 5.14\% \\
 &  N@10 & 0.4083 & \textbf{0.4115} & 0.78\% & 0.4010 & \textbf{0.4052} & 1.05\% & \textbf{0.4222} & 0.4215 & - & 0.1737 & \textbf{0.1775} & 2.19\% \\
 &  HR@5 & 0.5088 & \textbf{0.5157} & 1.36\% & 0.5003 & \textbf{0.5076} & 1.46\% & 0.5291 & \textbf{0.5329} & 0.72\% & 0.2046 & \textbf{0.2103} & 2.79\% \\
 &  HR@10 & 0.6856 & \textbf{0.6887} & 0.45\% & 0.6816 & \textbf{0.6849} & 0.48\% & 0.6978 & \textbf{0.7031} & 0.76\% & \textbf{0.3462} & 0.3425 & - \\
\bottomrule
\end{tabular}
}

\caption{Effectiveness of \model{} in collaborative filtering. We bold the best results.}
\label{tab:cf}

\end{table*}

%% file: tables/ctr.tex
\begin{table*}[]
\centering
\renewcommand\arraystretch{1.15}

\resizebox{.875\linewidth}{!}{
\begin{tabular}{cc|ccc|ccc|ccc|ccc}
\toprule
\multicolumn{2}{c|}{\textbf{Models}}
& \makecell{\textbf{DCN} \\ \small{(\citeyear{dcn})}}
& \makecell{\textbf{\model{}}$_\text{DCN}$ \\ \small{(ours)}}
& \textbf{Imp.}
& \makecell{\textbf{DeepFM} \\ \small{(\citeyear{deepfm})}}
& \makecell{\textbf{\model{}}$_\text{DeepFM}$ \\ \small{(ours)}}
& \textbf{Imp.}
& \makecell{\textbf{FiBiNET} \\ \small{(\citeyear{fibinet})}}
& \makecell{\textbf{\model{}}$_\text{FiBiNET}$ \\ \small{(ours)}}
& \textbf{Imp.}
& \makecell{\textbf{FinalMLP} \\ \small{(\citeyear{mao2023finalmlp})}}
& \makecell{\textbf{\model{}}$_\text{FinalMLP}$ \\ \small{(ours)}}
& \textbf{Imp.}
 \\ 
\midrule

\multirow{4}{*}{
\rotatebox{90}{\textbf{MIND}}
% \textbf{MIND}
}
 & \textbf{N@5}
    & 0.2031 & \textbf{0.2173} & 6.99\%
    & 0.2170 & \textbf{0.2428} & 11.89\%
    & 0.2181 & \textbf{0.2319} & 6.33\%
    & 0.2176 & \textbf{0.2265} & 4.09\%
    \\
 & \textbf{N@10} 
    & 0.2623 & \textbf{0.2770} & 5.60\%
    & 0.2749 & \textbf{0.2992} & 8.83\%
    & 0.2760 & \textbf{0.2881} & 4.38\%
    & 0.2757 & \textbf{0.2823} & 2.39\%
    \\
 & \textbf{HR@5}    
    & 0.3958 & \textbf{0.4101} & 3.61\%
    & 0.4065 & \textbf{0.4353} & 7.08\%
    & 0.4081 & \textbf{0.4336} & 6.25\%
    & 0.4061 & \textbf{0.4202} & 3.47\%
    \\
 & \textbf{HR@10}   
    & 0.5889 & \textbf{0.6012} & 2.09\%
    & 0.5948 & \textbf{0.6156} & 3.49\%
    & 0.5949 & \textbf{0.6129} & 3.03\%
    & 0.5935 & \textbf{0.6007} & 1.21\%
    \\
\midrule
\multirow{4}{*}{
\rotatebox{90}{\textbf{MovieLens}}
% \textbf{MovieLens}
}
 & \textbf{N@1}
    & 0.6781 & \textbf{0.7047} & 3.92\%
    & 0.6640 & \textbf{0.7035} & 5.95\%
    & 0.7016 & \textbf{0.7328} & 4.45\%
    & 0.7000 & \textbf{0.7440} & 6.29\%
    \\
 & \textbf{N@5} 
    & 0.7029 & \textbf{0.7204} & 2.49\%
    & 0.7014 & \textbf{0.7152} & 1.97\% 
    & 0.7314 & \textbf{0.7445} & 1.79\%
    & 0.7337 & \textbf{0.7431} & 1.28\%
    \\
 & \textbf{N@10}    
    & 0.7465 & \textbf{0.7569} & 1.39\%
    & 0.7433 & \textbf{0.7524} & 1.22\%
    & 0.7679 & \textbf{0.7862} & 2.38\%
    & 0.7696 & \textbf{0.7818} & 1.59\%
    \\
 & \textbf{HR@5}   
    & 0.9969 & \textbf{0.9984} & 0.15\%
    & 0.9969 & \textbf{0.9984} & 0.15\%
    & 0.9953 & \textbf{0.9969} & 0.16\%
    & 0.9937 & \textbf{0.9987} & 0.50\%
    \\
\bottomrule
\end{tabular}
}

\caption{Effectiveness of \model{} in click-through rate prediction. We bold the best results.}
\label{tab:ctr}

\end{table*}

%% file: tables/layers.tex
\begin{table*}[h]
\small
\centering
\renewcommand\arraystretch{1.15}

\resizebox{.9\linewidth}{!}{
\begin{tabular}{ccc|cccc|cccc|cccc|cccc}
\toprule
\multicolumn{3}{c|}{\textbf{Datasets}} & \multicolumn{8}{c|}{\textbf{Zhihu}} & \multicolumn{8}{c}{\textbf{Goodreads}} \\
\cmidrule(lr){1-3} \cmidrule(lr){4-11} \cmidrule(lr){12-19}
\multicolumn{3}{c|}{\textbf{Models}} & \multicolumn{4}{c|}{\textbf{\model{}}$_\text{BERT4Rec}$} & \multicolumn{4}{c|}{\textbf{\model{}}$_\text{FANS}$} & \multicolumn{4}{c|}{\textbf{\model{}}$_\text{BERT4Rec}$} & \multicolumn{4}{c}{\textbf{\model{}}$_\text{FANS}$} \\
\cmidrule(lr){1-3} \cmidrule(lr){4-7} \cmidrule(lr){8-11} \cmidrule(lr){12-15} \cmidrule(lr){16-19}
\textbf{$v^1$} & \textbf{$v^2$} & \textbf{$v^3$} & \textbf{N@5} & \textbf{N@10} & \textbf{HR@5} & \textbf{HR@10} & \textbf{N@5} & \textbf{N@10} & \textbf{HR@5} & \textbf{HR@10} & \textbf{N@5} & \textbf{N@10} & \textbf{HR@5} & \textbf{HR@10} & \textbf{N@5} & \textbf{N@10} & \textbf{HR@5} & \textbf{HR@10} \\
\midrule
\textbf{10} & - & -
    & 0.0197 & \underline{0.0295} & 0.2151 & 0.3752 
    & 0.0280 & \underline{0.0412} & \underline{0.3002} & \underline{0.4839}
    & \underline{0.0113} & \underline{0.0165} & \underline{0.1654} & 0.2510 
    & \textbf{0.0396} & \textbf{0.0561} & \textbf{0.4267} & \textbf{0.5473} \\
\textbf{20} & - & -
    & \textbf{0.0208} & \textbf{0.0307} & \textbf{0.2327} & \textbf{0.3953} 
    & 0.0273 & 0.0403 & 0.2895 & 0.4745 
    & 0.0105 & 0.0151 & \underline{0.1654} & \underline{0.2588}
    & \underline{0.0388} & \underline{0.0554} & \underline{0.4120} & \underline{0.5363} \\
\textbf{50} & - & -
    & \underline{0.0199} & 0.0277 & \underline{0.2204} & \underline{0.3852} 
    & \textbf{0.0299} & 0.0411 & 0.2975 & 0.4758 
    & \textbf{0.0119} & \textbf{0.0169} & \textbf{0.1673} & 0.2549 
    & 0.0375 & 0.0540 & 0.4066 & 0.5288\\
\textbf{100} & - & -
    & 0.0142 & 0.0214 & 0.1691 & 0.3081 
    & \underline{0.0288} & \textbf{0.0414} & \textbf{0.3044} & \textbf{0.4866} 
    & 0.0104 & 0.0153 & 0.1511 & 0.2412 
    & 0.0366 & 0.0510 & 0.4034 & 0.5201 \\
\textbf{500} & - & -
    & 0.0057 & 0.0094 & 0.0910 & 0.1752 
    & 0.0249 & 0.0397 & 0.2702 & 0.4685
    & 0.0110 & 0.0158 & \underline{0.1654} & \textbf{0.2601}
    & 0.0367 & 0.0516 & 0.3865 & 0.5071 \\
\midrule
\textbf{100} & \textbf{10} & -
    & 0.0192 & 0.0271 & 0.2129 & 0.3705 
    & 0.0270 & 0.0384 & 0.2884 & 0.4651 
    & 0.0096 & 0.0148 & 0.1654 & 0.2588 
    & \underline{0.0387} & \underline{0.0555} & 0.4021 & \underline{0.5383} \\
\textbf{200} & \textbf{10} & -
    & \textbf{0.0235} & \textbf{0.0311} & \textbf{0.2429} & 0.3832 
    & 0.0271 & 0.0389 & 0.2970 & 0.4705 
    & \textbf{0.0130} & \textbf{0.0180} & \textbf{0.1829} & 0.2678 
    & 0.0368 & 0.0531 & 0.3872 & 0.5350 \\
\textbf{500} & \textbf{10} & -
    & \underline{0.0223} & 0.0307 & \underline{0.2376} & 0.3966
    & \textbf{0.0301} & \textbf{0.0428} & \underline{0.3034} & \underline{0.4859}
    & 0.0108 & 0.0163 & 0.1673 & \underline{0.2724} 
    & 0.0379 & 0.0540 & \underline{0.4092} & 0.5363 \\ 
\textbf{1000} & \textbf{10} & -
    & 0.0219 & \underline{0.0308} & 0.2322 & \underline{0.4040}
    & 0.0251 & 0.0364 & 0.2627 & 0.4436 
    & 0.0084 & 0.0119 & 0.1386 & 0.2121 
    & 0.0344 & 0.0491 & 0.3761 & 0.5032 \\
\textbf{200} & \textbf{20} & -
    & 0.0165 & 0.0244 & 0.1851 & 0.3275
    & 0.0282 & 0.0412 & \textbf{0.3066} & 0.4899
    & \underline{0.0116} & 0.0159 & 0.1699 & 0.2536
    & 0.0379 & 0.0542 & 0.3988 & 0.5318 \\
\textbf{400} & \textbf{20} & -
    & 0.0182 & 0.0276 & 0.2113 & 0.3799
    & 0.0279 & 0.0412 & 0.2884 & \textbf{0.4919} 
    & 0.0088 & 0.0133 & 0.1388 & 0.2348
    & 0.0368 & 0.0538 & 0.3956 & 0.5337 \\
\textbf{8000} & \textbf{20} & -
    & 0.0197 & 0.0291 & 0.2349 & \textbf{0.4047}
    & 0.0268 & 0.0395 & 0.2900 & 0.4826 
    & 0.0089 & 0.0125 & 0.1420 & 0.2185 
    & 0.0373 & 0.0539 & \underline{0.4092} & \underline{0.5383} \\
\textbf{500} & \textbf{50} & -
    & 0.0114 & 0.0164 & 0.1482 & 0.2631
    & \underline{0.0294} & \underline{0.0419} & 0.3028 & 0.4893
    & 0.0112 & \underline{0.0166} & \underline{0.1783} & \textbf{0.2769}
    & \textbf{0.0399} & \textbf{0.0567} & \textbf{0.4275} & \textbf{0.5473} \\
\textbf{2500} & \textbf{50} & -
    & 0.0095 & 0.0148 & 0.1348 & 0.2517
    & 0.0264 & 0.0388 & 0.2868 & 0.4631
    & 0.0103 & 0.0148 & 0.1556 & 0.2510
    & 0.0359 & 0.0516 & 0.3761 & 0.5065 \\
\midrule
\textbf{4000} & \textbf{200} & \textbf{10}
    & \underline{0.0199} & \textbf{0.0294} & \textbf{0.2301} & \textbf{0.3906} 
    & \textbf{0.0266} & \textbf{0.0388} & \textbf{0.2895} & \textbf{0.4765} 
    & \textbf{0.0090} & \underline{0.0134} & \textbf{0.1446} & \textbf{0.2425}
    & \textbf{0.0393} & \textbf{0.0553} & \textbf{0.4040} & \textbf{0.5435} \\
\textbf{8000} & \textbf{400} & \textbf{20}
    & \textbf{0.0200} & \underline{0.0285} & \underline{0.2204} & \underline{0.3691} 
    & \underline{0.0245} & \underline{0.0365} & \underline{0.2755} & \underline{0.4651} 
    & \underline{0.0088} & \textbf{0.0136} & \underline{0.1381} & \textbf{0.2425}
    & \underline{0.0383} & \underline{0.0544} & \underline{0.4008} & \underline{0.5350} \\
\bottomrule
\end{tabular}
}

\caption{Impact of the number of \model{} layers (H) and the number of entries of each layer ($v^{i}$). The best results are indicated in bold, while the second-best results are underlined. A hyphen (-) indicates the absence of a layer. For example, ``100($v^{1}$) 10($v^{2}$) -($V^{3}$)'' means that \model{} only has two layers, and the first and second layers correspond to the 100-entry and 10-entry codebooks, respectively. We fix $\alpha, \beta, \omega_\text{c}, \omega_\text{q}$ to be $1.0$ in this experiment.}
\label{tab:layers}
\end{table*}

%% file: plots/dual/main.tex
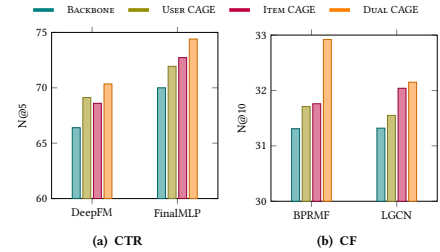
\begin{figure}
    \centering
    \setlength\tabcolsep{2pt}
    \resizebox{.7\linewidth}{!}{
    \begin{tabular}{cc}
    \multicolumn{2}{c}{
        \resizebox{1.0\linewidth}{!}{
            \input{plots/dual/legend}
        }
    } \\
    \begin{subfigure}{0.25\textwidth}
        \resizebox{1.0\linewidth}{!}{
            \input{plots/dual/ctr}
        }
        \caption{\label{fig:dual-ctr} CTR}
    \end{subfigure} &
    \begin{subfigure}{0.25\textwidth}
        \resizebox{1.0\linewidth}{!}{
            \input{plots/dual/cf}
        }
        \caption{\label{fig:dual-cf} CF}
    \end{subfigure}
    \end{tabular}
    }
    
    \caption{\label{fig:dual}Influence of the use of user and item \model{} in the non-sequential recommenders.}
\end{figure}

%% file: plots/dual/legend.tex
\begin{tikzpicture}
    \begin{customlegend}[
        legend columns=4,
        legend style={
            align=left,
            draw=none,
            column sep=2ex
        },
        legend entries={
            \textsc{Backbone},
            \textsc{User \model{}},
            \textsc{Item \model{}},
            \textsc{Dual \model{}},
        }]
        % \addlegendimage{teal,mark=x,solid,fill=teal!30}
        % \addlegendimage{purple,mark=x,solid,fill=purple!30}
        % \addlegendimage{olive,mark=x,solid,fill=olive!30}
        % \addlegendimage{orange,mark=x,solid,fill=orange!30}
\addlegendimage{teal,mark=none,draw=teal,fill=teal!30,line width=3pt}
\addlegendimage{olive,mark=none,draw=olive,fill=olive!30,line width=3pt}
\addlegendimage{purple,mark=none,draw=purple,fill=purple!30,line width=3pt}
\addlegendimage{orange,mark=none,draw=orange,fill=orange!30,line width=3pt}
        \end{customlegend}
\end{tikzpicture}

%% file: plots/dual/ctr.tex
\begin{tikzpicture}
\begin{axis}[
    ybar,
    ymin=60,
    ymax=75,
    xmin=0,
    xmax=3,
    xtick=data,
    xticklabels={DeepFM, FinalMLP},
    ylabel=N@5,
    bar width=6pt,
%    xlabel=Models,
    enlarge x limits=0.15,
    width=6cm,
    height=6cm,
]
\addplot[teal!80!black, fill=teal!50!white] coordinates {(0.5,66.40) (2.5,70.00)};
\addplot[olive!80!black, fill=olive!50!white] coordinates {(0.5,69.12) (2.5,71.94)};
\addplot[purple!80!black, fill=purple!50!white] coordinates {(0.5,68.60) (2.5,72.73)};
\addplot[orange!80!black, fill=orange!50!white] coordinates {(0.5,70.35) (2.5,74.40)};
\end{axis}
\end{tikzpicture}

%% file: plots/dual/cf.tex
\begin{tikzpicture}
\begin{axis}[
    ybar,
    ymin=30,
    ymax=33,
    xmin=0,
    xmax=3,
    xtick=data,
    xticklabels={BPRMF, LGCN},
    ylabel=N@10,
    bar width=6pt,
%    xlabel=Models,
    enlarge x limits=0.15,
    width=6cm,
    height=6cm,
]
\addplot[teal!80!black, fill=teal!50!white] coordinates {(0.5,31.31) (2.5,31.32)};
\addplot[olive!80!black, fill=olive!50!white] coordinates {(0.5,31.71) (2.5,31.55)};
\addplot[purple!80!black, fill=purple!50!white] coordinates {(0.5,31.76) (2.5,32.04)};
\addplot[orange!80!black, fill=orange!50!white] coordinates {(0.5,32.92) (2.5,32.15)};
\end{axis}
\end{tikzpicture}

%% file: plots/hyper.tex
\begin{figure*}[h]
    \centering
    \setlength\tabcolsep{2pt}
    \resizebox{.75\linewidth}{!}{
    \begin{tabular}{cccc}
    \multicolumn{4}{c}{
        % \resizebox{1.0\linewidth}{!}{
            \input{plots/alpha/legend}
        % }
    } \\
    
    \begin{subfigure}{0.25\textwidth}
        \resizebox{1.0\linewidth}{!}{
            \input{plots/alpha/zhihu}
        }
        \caption{\label{fig:alpha-zhihu}Zhihu}
    \end{subfigure} &
    
    \begin{subfigure}{0.25\textwidth}
        \resizebox{1.0\linewidth}{!}{
            \input{plots/beta/zhihu}
        }
        \caption{\label{fig:beta-zhihu}Zhihu}
    \end{subfigure} &    
    \begin{subfigure}{0.25\textwidth}
        \resizebox{1.0\linewidth}{!}{
            \input{plots/omega_code/zhihu}
        }
        \caption{\label{fig:omega_code-zhihu}Zhihu}
    \end{subfigure} &
    
    \begin{subfigure}{0.25\textwidth}
        \resizebox{1.0\linewidth}{!}{
            \input{plots/omega_quant/zhihu}
        }
        \caption{\label{fig:omega_quant-zhihu}Zhihu}
    \end{subfigure} \\

    \begin{subfigure}{0.25\textwidth}
        \resizebox{1.0\linewidth}{!}{
            \input{plots/alpha/goodreads}
        }
        \caption{\label{fig:alpha-goodreads}Goodreads}
    \end{subfigure} &
    
    \begin{subfigure}{0.25\textwidth}
        \resizebox{1.0\linewidth}{!}{
            \input{plots/beta/goodreads}
        }
        \caption{\label{fig:beta-goodreads}Goodreads}
    \end{subfigure} &
    
    \begin{subfigure}{0.25\textwidth}
        \resizebox{1.0\linewidth}{!}{
            \input{plots/omega_code/goodreads}
        }
        \caption{\label{fig:omega_code-goodreads}Goodreads}
    \end{subfigure} &
    
    \begin{subfigure}{0.25\textwidth}
        \resizebox{1.0\linewidth}{!}{
            \input{plots/omega_quant/goodreads}
        }
        \caption{\label{fig:omega_quant-goodreads}Goodreads}
    \end{subfigure} \\
    \end{tabular}
    }
    
    \caption{\label{fig:hyper}
    Impact of the residual connection weight $\alpha$, the quantization commitment cost $\beta$, the codebook classification loss weight $\omega_\text{c}$, and the quantization loss weight $\omega_\text{q}$.
    We use the model with $\alpha=0$ as the reference baseline for (a), and measure the \textit{relative improvement} of each metric compared to the baseline for various values of $\alpha$, defined as $(m_\alpha - m_\text{0}) / m_\text{0} * 100\%$, where $m$ is one of the metrics in \{N@5, N@10, HR@5, HR@10\}.
    Therefore, the relative improvement of $\alpha=0$ is constant at 0\%.
    Similarly, we use the model with $\beta=0$ as the reference baseline for (b), $\omega_\text{q}=0$ for (c), and $\omega_\text{q}=0$ for (d).
    % All the experiments are conducted on the Zhihu dataset.
    }
\end{figure*}
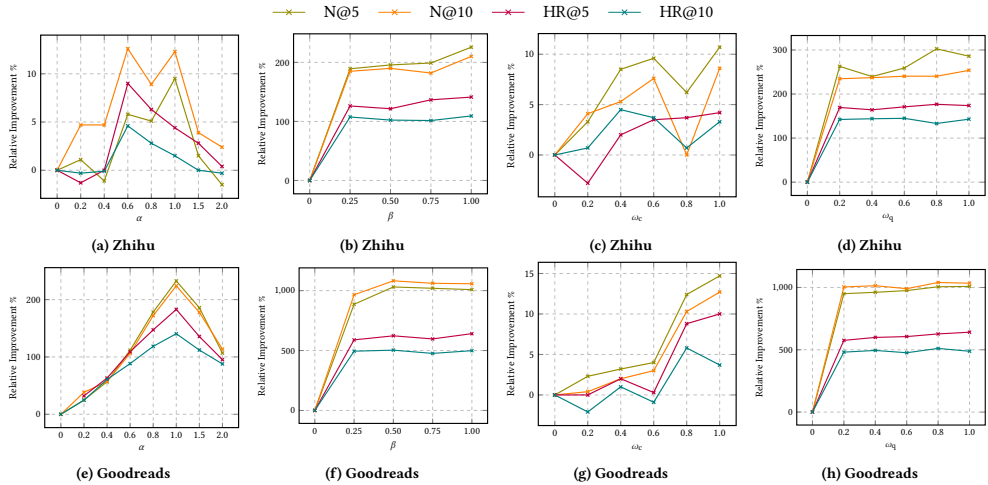

%% file: plots/alpha/legend.tex
\begin{tikzpicture}
    \begin{customlegend}[
        legend columns=4,
        legend style={
            align=left,
            draw=none,
            column sep=2ex
        },
        legend entries={
            \textsc{N@5},
            \textsc{N@10},
            \textsc{HR@5},
            \textsc{HR@10},
        }]
        \addlegendimage{olive,mark=x,solid,line legend}
        \addlegendimage{orange,mark=x,solid,line legend}
        \addlegendimage{purple,mark=x,solid,line legend}
        \addlegendimage{teal,mark=x,solid,line legend}
        \end{customlegend}
\end{tikzpicture}

%% file: plots/alpha/zhihu.tex
\begin{tikzpicture}
    \begin{axis}[
        ylabel=Relative Improvement \%,
        xlabel=$\alpha$,
        xtick={0,1,2,3,4,5,6,7,8},
        xticklabels={0, 0.2, 0.4, 0.6, 0.8, 1.0, 1.5, 2.0},
        line width=0.35mm,
        grid=both,
        grid style=dashed,
    ]

        \addplot[
            olive,
            mark=x,
        ]
            coordinates{
                (0, 0)
                (1, 1.1)
                (2, -1.1)
                (3, 5.8)
                (4, 5.1)
                (5, 9.5)
                (6, 1.5)
                (7, -1.5)
            }; \label{plot:alpha-zhihu-n5}
            
        \addplot[
            orange,
            mark=x,
        ]
            coordinates{
                (0, 0)
                (1, 4.7)
                (2, 4.7)
                (3, 12.6)
                (4, 8.9)
                (5, 12.3)
                (6, 3.9)
                (7, 2.4)
            }; \label{plot:alpha-zhihu-n10}

        \addplot[
            purple,
            mark=x,
        ]
            coordinates{
                (0, 0)
                (1, -1.3)
                (2, 0.0)
                (3, 9.0)
                (4, 6.3)
                (5, 4.4)
                (6, 2.8)
                (7, 0.4)
            }; \label{plot:alpha-zhihu-hr5}

        \addplot[
            teal,
            mark=x,
        ]
            coordinates{
                (0, 0)
                (1, -0.3)
                (2, -0.1)
                (3, 4.6)
                (4, 2.8)
                (5, 1.5)
                (6, 0.0)
                (7, -0.3)
            }; \label{plot:alpha-zhihu-hr10}
    \end{axis}
\end{tikzpicture}

%% file: plots/beta/zhihu.tex
\begin{tikzpicture}
    \begin{axis}[
        ylabel=Relative Improvement \%,
        xlabel=$\beta$,
        xtick={0,1,2,3,4},
        xticklabels={0, 0.25, 0.50, 0.75, 1.00},
        line width=0.35mm,
        grid=both,
        grid style=dashed,
    ]

        \addplot[
            olive,
            mark=x,
        ]
            coordinates{
                (0, 0)
                (1, 189.2)
                (2, 195.7)
                (3, 198.9)
                (4, 225.8)
            }; \label{plot:beta-zhihu-n5}
            
        \addplot[
            orange,
            mark=x,
        ]
            coordinates{
                (0, 0)
                (1, 184.9)
                (2, 189.9)
                (3, 182.0)
                (4, 210.1)
            }; \label{plot:beta-zhihu-n10}

        \addplot[
            purple,
            mark=x,
        ]
            coordinates{
                (0, 0)
                (1, 126.2)
                (2, 121.5)
                (3, 136.6)
                (4, 141.2)
            }; \label{plot:beta-zhihu-hr5}

        \addplot[
            teal,
            mark=x,
        ]
            coordinates{
                (0, 0)
                (1, 107.5)
                (2, 102.3)
                (3, 101.5)
                (4, 109.3)
            }; \label{plot:beta-zhihu-hr10}
    \end{axis}
\end{tikzpicture}

%% file: plots/omega_code/zhihu.tex
\begin{tikzpicture}
    \begin{axis}[
        ylabel=Relative Improvement \%,
        xlabel=$\omega_\text{c}$,
        xtick={0,1,2,3,4,5,6},
        xticklabels={0, 0.2, 0.4, 0.6, 0.8, 1.0},
        line width=0.35mm,
        grid=both,
        grid style=dashed,
    ]

        \addplot[
            olive,
            mark=x,
        ]
            coordinates{
                (0, 0)
                (1, 3.3)
                (2, 8.5)
                (3, 9.6)
                (4, 6.2)
                (5, 10.7)
            }; \label{plot:omega_code-zhihu-n5}
            
        \addplot[
            orange,
            mark=x,
        ]
            coordinates{
                (0, 0)
                (1, 4.1)
                (2, 5.3)
                (3, 7.6)
                (4, 0.0)
                (5, 8.6)
            }; \label{plot:omega_code-zhihu-n10}

        \addplot[
            purple,
            mark=x,
        ]
            coordinates{
                (0, 0)
                (1, -2.8)
                (2, 2.0)
                (3, 3.5)
                (4, 3.7)
                (5, 4.2)
            }; \label{plot:omega_code-zhihu-hr5}

        \addplot[
            teal,
            mark=x,
        ]
            coordinates{
                (0, 0)
                (1, 0.7)
                (2, 4.5)
                (3, 3.7)
                (4, 0.7)
                (5, 3.3)
            }; \label{plot:omega_code-zhihu-hr10}
    \end{axis}
\end{tikzpicture}

%% file: plots/omega_quant/zhihu.tex
\begin{tikzpicture}
    \begin{axis}[
        ylabel=Relative Improvement \%,
        xlabel=$\omega_\text{q}$,
        xtick={0,1,2,3,4,5,6},
        xticklabels={0, 0.2, 0.4, 0.6, 0.8, 1.0},
        line width=0.35mm,
        grid=both,
        grid style=dashed,
    ]

        \addplot[
            olive,
            mark=x,
        ]
            coordinates{
                (0, 0)
                (1, 262.8)
                (2, 239.7)
                (3, 259.0)
                (4, 302.6)
                (5, 285.9)
            }; \label{plot:omega_quant-zhihu-n5}
            
        \addplot[
            orange,
            mark=x,
        ]
            coordinates{
                (0, 0)
                (1, 234.7)
                (2, 237.2)
                (3, 240.5)
                (4, 240.5)
                (5, 253.7)
            }; \label{plot:omega_quant-zhihu-n10}

        \addplot[
            purple,
            mark=x,
        ]
            coordinates{
                (0, 0)
                (1, 169.5)
                (2, 164.2)
                (3, 170.9)
                (4, 176.7)
                (5, 173.8)
            }; \label{plot:omega_quant-zhihu-hr5}

        \addplot[
            teal,
            mark=x,
        ]
            coordinates{
                (0, 0)
                (1, 142.3)
                (2, 143.9)
                (3, 144.9)
                (4, 133.2)
                (5, 142.9)
            }; \label{plot:omega_quant-zhihu-hr10}
    \end{axis}
\end{tikzpicture}

%% file: plots/alpha/goodreads.tex
\begin{tikzpicture}
    \begin{axis}[
        ylabel=Relative Improvement \%,
        xlabel=$\alpha$,
        xtick={0,1,2,3,4,5,6,7,8},
        xticklabels={0, 0.2, 0.4, 0.6, 0.8, 1.0, 1.5, 2.0},
        line width=0.35mm,
        grid=both,
        grid style=dashed,
    ]

        \addplot[
            olive,
            mark=x,
        ]
            coordinates{
                (0, 0)
                (1, 25.0)
                (2, 56.7)
                (3, 111.7)
                (4, 178.3)
                (5, 232.5)
                (6, 185.8)
                (7, 106.7)
            }; \label{plot:alpha-goodreads-n5}
            
        \addplot[
            orange,
            mark=x,
        ]
            coordinates{
                (0, 0)
                (1, 38.3)
                (2, 57.1)
                (3, 106.3)
                (4, 172.0)
                (5, 224.0)
                (6, 177.7)
                (7, 113.7)  
            }; \label{plot:alpha-goodreads-n10}

        \addplot[
            purple,
            mark=x,
        ]
            coordinates{
                (1, 32.2)
                (2, 63.1)
                (3, 109.0)
                (4, 147.2)
                (5, 182.9)
                (6, 135.6)
                (7, 95.3)
            }; \label{plot:alpha-goodreads-hr5}

        \addplot[
            teal,
            mark=x,
        ]
            coordinates{
                (0, 0)
                (1, 24.5)
                (2, 60.7)
                (3, 88.4)
                (4, 118.5)
                (5, 140.5)
                (6, 112.0)
                (7, 87.8)
            }; \label{plot:alpha-goodreads-hr10}
    \end{axis}
\end{tikzpicture}

%% file: plots/beta/goodreads.tex
\begin{tikzpicture}
    \begin{axis}[
        ylabel=Relative Improvement \%,
        xlabel=$\beta$,
        xtick={0,1,2,3,4},
        xticklabels={0, 0.25, 0.50, 0.75, 1.00},
        line width=0.35mm,
        grid=both,
        grid style=dashed,
    ]

        \addplot[
            olive,
            mark=x,
        ]
            coordinates{
                (0, 0)
                (1, 886.1)
                (2, 1030.6)
                (3, 1019.4)
                (4, 1008.3)
            }; \label{plot:beta-goodreads-n5}
            
        \addplot[
            orange,
            mark=x,
        ]
            coordinates{
                (0, 0)
                (1, 965.3)
                (2, 1081.6)
                (3, 1061.2)
                (4, 1057.1)
            }; \label{plot:beta-goodreads-n10}

        \addplot[
            purple,
            mark=x,
        ]
            coordinates{
                (0, 0)
                (1, 588.9)
                (2, 622.7)
                (3, 596.9)
                (4, 640.9)
            }; \label{plot:beta-goodreads-hr5}

        \addplot[
            teal,
            mark=x,
        ]
            coordinates{
                (0, 0)
                (1, 494.6)
                (2, 503.1)
                (3, 476.1)
                (4, 498.8)
            }; \label{plot:beta-goodreads-hr10}
    \end{axis}
\end{tikzpicture}

%% file: plots/omega_code/goodreads.tex
\begin{tikzpicture}
    \begin{axis}[
        ylabel=Relative Improvement \%,
        xlabel=$\omega_\text{c}$,
        xtick={0,1,2,3,4,5,6},
        xticklabels={0, 0.2, 0.4, 0.6, 0.8, 1.0},
        line width=0.35mm,
        grid=both,
        grid style=dashed,
    ]

        \addplot[
            olive,
            mark=x,
        ]
            coordinates{
                (0, 0)
                (1, 2.3)
                (2, 3.2)
                (3, 4.0)
                (4, 12.4)
                (5, 14.7)
            }; \label{plot:omega_code-goodreads-n5}
            
        \addplot[
            orange,
            mark=x,
        ]
            coordinates{
                (0, 0)
                (1, 0.4)
                (2, 2.0)
                (3, 3.0)
                (4, 10.3)
                (5, 12.7)
            }; \label{plot:omega_code-goodreads-n10}

        \addplot[
            purple,
            mark=x,
        ]
            coordinates{
                (0, 0)
                (1, 0.0)
                (2, 2.0)
                (3, 0.3)
                (4, 8.8)
                (5, 10.0)
            }; \label{plot:omega_code-goodreads-hr5}

        \addplot[
            teal,
            mark=x,
        ]
            coordinates{
                (0, 0)
                (1, -2.1)
                (2, 1.0)
                (3, -0.9)
                (4, 5.8)
                (5, 3.7)
            }; \label{plot:omega_code-goodreads-hr10}
    \end{axis}
\end{tikzpicture}

%% file: plots/omega_quant/goodreads.tex
\begin{tikzpicture}
    \begin{axis}[
        ylabel=Relative Improvement \%,
        xlabel=$\omega_\text{q}$,
        xtick={0,1,2,3,4,5,6},
        xticklabels={0, 0.2, 0.4, 0.6, 0.8, 1.0},
        line width=0.35mm,
        grid=both,
        grid style=dashed,
    ]

        \addplot[
            olive,
            mark=x,
        ]
            coordinates{
                (0, 0)
                (1, 950.0)
                (2, 961.1)
                (3, 975.0)
                (4, 1005.6)
                (5, 1008.3)
            }; \label{plot:omega_code-goodreads-n5}
            
        \addplot[
            orange,
            mark=x,
        ]
            coordinates{
                (0, 0)
                (1, 1004.0)
                (2, 1014.0)
                (3, 990.0)
                (4, 1040.0)
                (5, 1034.0)
            }; \label{plot:omega_code-goodreads-n10}

        \addplot[
            purple,
            mark=x,
        ]
            coordinates{
                (0, 0)
                (1, 575.6)
                (2, 599.1)
                (3, 605.9)
                (4, 627.2)
                (5, 640.9)
            }; \label{plot:omega_code-goodreads-hr5}

        \addplot[
            teal,
            mark=x,
        ]
            coordinates{
                (0, 0)
                (1, 481.1)
                (2, 494.6)
                (3, 476.1)
                (4, 510.9)
                (5, 487.9)
            }; \label{plot:omega_code-goodreads-hr10}
    \end{axis}
\end{tikzpicture}

%% file: content/Conclusion.tex
\section{Conclusion}

%In this paper, we propose a novel approach, namely \model{}, to enhance the category-awareness and robustness of item representations in id-based recommender systems. Our approach is based on the vector quantization technique, which is capable of learning hierarchical categorical knowledge at different levels of granularity, and can be easily integrated into a wide range of existing recommenders of various types. Through extensive experiments conducted on diverse recommendation scenarios and various real-world datasets, we have demonstrated the effectiveness of \model{} in improving the performance of baseline recommendation models. Our visualization experiments also indicate the robustness of the category-aware code representation.

We have proposed \model{}, a novel framework for leaning item/user category trees for ID-based recommendation, by employing differentiable vector quantization techniques.
%By using vector quantization techniques, \model{} enables the category learning at multiple granularity levels. 
The flexibility of \model{} allows for its seamless integration into a variety of existing recommender systems. 
Through comprehensive experiments conducted across diverse recommendation scenarios, 
we have demonstrated the effectiveness of \model{} in enhancing the performance of various recommendation models. In future work, we plan to explore extending the applicability of CAGE beyond ID-based recommendation.

%In the future, we would like to study its applicability beyond ID-based recommendation.

\section*{Acknowledgments}
We thank the anonymous reviewers
% and meta review 
for their valuable feedback. 

%% file: content/Appendix.tex
\section{Data Statistics}\label{sec:statistics}

The dataset statistics are summarized in Table~\ref{tab:statistics}.

\section{Implementation Details}

In our training process, we employ the Adam optimizer as the gradient descent algorithm. Across all models, the embedding dimension is consistently set to 64.

\subsection{List Completion}

We adhere to the experimental setup established by FANS~\cite{fans} and maintain the following parameters:

For all attention-based models (SASRec, BERT4Rec, and FANS), we set the number of hidden layers as 3, and the number of attention heads as 8. For the Caser model, we specify the number of horizontal and vertical channels as 16 and 4, respectively, and set the max sequence length to 5. For the CAR model, we define the number of CAR layers as 3. As for the GRU4Rec model, we use 3 GRU layers.

For the BERT4Rec model, each item in the sequence has a 
$S$ probability to be selected. For each selected item, it has 1) an 80\% probability to be replaced with a $<MASK>$ token; 2) a 10\% probability for substitution with a random item, and 3) the remaining 10\% probability to remain unchanged. Furthermore, there is an additional probability of $A$, resulting in a combined probability of $S+A$ for the last token to be masked. This strategy is proposed by BERT4Rec for the sequential recommendation scenario that requires next item prediction during inference. In our experiments, we vary the value of \textbf{S} from the set $\{15, 30, 45, 60\}$ while keeping 
$A$ fixed at 10. Conversely, we also vary the value of $A$ from the set $\{10, 20, 30, 40\}$ while keeping $S$ fixed at 15. Finally, we select the configuration of $S=15$ and $A=10$ for both the original and CAGE-integrated models.

Moreover, for all the aforementioned baseline models, the batch size is set to 256, and we explore a range of learning rates from $\{0.001, 0.005, 0.01, 0.05, 0.1\}$.

\subsection{Collaborative Filtering}

In this scenario, we adhere to the default setting of ReChorus~\cite{wang2020make}, which serves as the foundation for our implementation. Subsequently, based on the default setting, we execute a comprehensive grid search of significant hyperparameters. These include the learning rate, which is selected from $\{0.05, 0.001, 0.005, 1e-4, 1e-5, 1e-6\}$, and L2 that ranges from $1e-8$ to $1e-5$ using a logarithmic scale.

Specifically, for the LightGCN model, we set the number of layers as 3. 
For the NeuMF model, we tune the number of layers from $\{1, 2, 3\}$. In the case of CFKG, we incorporate four types of relations: \textit{\textbf{user-item}: buy}, 
\textit{\textbf{item-category}: has\_cat}, 
\textit{\textbf{item-item}: also\_bought (complement)}, 
\textit{\textbf{item-item}: also\_view (substitute)}. 
Notably, the user-item matrix is incorporated into the knowledge graph as a unique relation, denoted as \textit{``buy''}.

\subsection{Click-through Rate Prediction}

For the CTR scenario, we set the batch size to 5000 and tune the learning rate from the set $\{0.0001, 0.0005, 0.001, 0.005, 0.01\}$. In addition, we maintain a uniform configuration across all base models, including setting the number of DNN layers to 3 and a dropout ratio of 0.1. For models like DCN, DeepFM, and FiBiNET, we vary the DNN hidden size from the set $\{500, 1000, 2000\}$. For the FinalMLP model, we tune the MLP hidden size from $\{64, 128, 256\}$. ReLU serves as the activation function for the DNN. As for the DCN model, we vary the number of cross layers from $\{3, 4, 5\}$.

\subsection{Integration of CAGE}

When integrating CAGE with base models, we tune the following parameters: 1) the number of CAGE layers, 2) the number of codes per layer, 3) the weight factor $\alpha$ for residual connections, 4) the quantization commitment cost $\beta$, 5) the codebook classification loss weight $\omega_c$ for the list completion task, and 6) the quantization loss weight $\omega_q$. For a comprehensive analysis of the impact of these parameters, please refer to the Abaltion Study section.

\section{Integrating CAGE with Other CTR Models}

We report additional experiments integrating our GAGE approach with other recent CTR models, i.e., MaskNet~\cite{wang2021masknet} and GDCN~\cite{wang2023towards}, on the MIND dataset, as illustrated in Table~\ref{tab:addition-ctr}, which have also showcased notable improvements.

\input{tables/statistics}

\begin{table}[]
\centering
\renewcommand\arraystretch{1.15}

\begin{tabular}{c|cccc}
\toprule
\textbf{Model} & \textbf{N@5} & \textbf{N@10} & \textbf{HR@5} & \textbf{HR@10} \\
\midrule
MaskNet & 0.2208 & 0.2796 & 0.4076 & 0.5936 \\
\model{}$_\text{MaskNet}$ & 0.2302 & 0.2872 & 0.4252 & 0.6093 \\
\midrule
GDCN & 0.2119 & 0.2762 & 0.3866 & 0.5942 \\
\model{}$_\text{GDCN}$ & 0.2273 & 0.2852 & 0.4160 & 0.6006 \\
\bottomrule
\end{tabular}

\caption{Integrating CAGE with other CTR models.}
\label{tab:addition-ctr}
\end{table}

\section{Deeper Analysis of the Categories Learned by CAGE}

We carry out further experiments on the MIND dataset, which comprises 270 real subcategories. We set the codebook size for the first and second layers as 1000 and 10, respectively. After training the \textbf{\model{}}$_\text{DeepFM}$ model, we examine the correlation between the learned codes of the first layer and the actual subcategories, specifically, how each category is distributed among the codes.

Out of the 270 real categories, 55 categories (over 20\%) are exclusive to one code. There are 116 categories (over 40\%) that are spread across fewer than 10 codes (1\%). Additionally, 148 categories (approximately 55\%) are distributed among fewer than 20 codes (2\%). Furthermore, 208 categories (over 75\%) are distributed among fewer than 100 codes (10\%). The results show a degree of correlation between the learned and actual categories, validating the effectiveness of \model{} in automatic categorization.

\begin{table}[]
\centering
\renewcommand\arraystretch{1.15}

\begin{tabular}{c|cc|cc}
\toprule
\textbf{DCN} & \textbf{Ori.} & \textbf{Ori. +} & \textbf{\model{}} & \textbf{\model{} +} \\
\midrule
\textbf{N@5} & 0.2031 & 0.2065 & 0.2173 & 0.2346 \\
\textbf{N@10} & 0.2623 & 0.2711 & 0.2770 & 0.2935 \\
\textbf{HR@5} & 0.3958 & 0.3880 & 0.4101 & 0.4229 \\
\textbf{HR@10} & 0.5889 & 0.5968 & 0.6012 & 0.6114 \\
\midrule
\textbf{FiBiNET} & \textbf{Ori.} & \textbf{Ori. +} & \textbf{\model{}} & \textbf{\model{} +} \\
\midrule
\textbf{N@5} & 0.2181 & 0.2212 & 0.2319 & 0.2492 \\
\textbf{N@10} & 0.2760 & 0.2808 & 0.2881 & 0.3065 \\
\textbf{HR@5} & 0.4081 & 0.4159 & 0.4336 & 0.4459 \\
\textbf{HR@10} & 0.5949 & 0.6065 & 0.6129 & 0.6253 \\
\bottomrule
\end{tabular}

\caption{Influence of the real categorical knowledge. ``Ori.'' represents the original base model. ``+'' represents the model that integrates category features.}
\label{tab:addition-cat}

\end{table}

\section{Integration of Real Category Data}

Here, we study the integration of the original baseline and \model{}-enhanced variants with the real category data on the MIND dataset. The category information is integrated as follows:
\begin{equation}
    \mathbf{e}^\prime = \mathbf{e} + \mathbf{e}_\text{cat},
\end{equation}
where $\mathbf{e}$ is the original item embedding, $\mathbf{e}_\text{cat}$ is the corresponding category embedding, and $\mathbf{e}^\prime$ is the category-informed item embedding. Such integration is denoted by ``+'' in Table~\ref{tab:addition-cat}. The results indicate that real category information can enhance both the baselines and our \model{} variants. Furthermore, it is surprising to note that our \model{} method outperforms ``Ori. +'' that utilizes real category information. A possible reason could be that \model{} acquires categorical information from interaction data, which might be more beneficial for the recommendation task. This could imply that the category knowledge learned by \model{} effectively complements the provided category information.

\begin{table}[]
\centering
\renewcommand\arraystretch{1.15}
\begin{tabular}{r|cccc}
\toprule
 & \textbf{N@5} & \textbf{N@10} & \textbf{HR@5} & \textbf{HR@10} \\
\midrule
\textbf{FinalMLP} & 0.2176 & 0.2757 & 0.4061 & 0.5935 \\
\textbf{\model{}}$_\text{FinalMLP}$ (avg) & 0.2265 & 0.2823 & 0.4202 & 0.6007 \\
\textbf{\model{}}$_\text{FinalMLP}$ (concat) & 0.2270 & 0.2872 & 0.4146 & 0.6060 \\
\midrule
\textbf{DCN} & 0.2031 & 0.2623 & 0.3958 & 0.5889 \\
\textbf{\model{}}$_\text{DCN}$ (avg) & 0.2173 & 0.2770 & 0.4101 & 0.6012 \\
\textbf{\model{}}$_\text{DCN}$ (concat) & 0.2196 & 0.2798 & 0.4020 & 0.5960 \\
\bottomrule
\end{tabular}
\end{table}

\section{Aggregation of Multi-level Category Embeddings}

Here, we explored different ways of aggregating the generated multi-level category embeddings, including average pooling, denoted as ``avg'', and concat-and-project (i.e., concatenate the multi-level category vectors into a long vector and then project it to the dimension of a single vector using a fully connected layer), denoted as ``concat''. Based on our experiments, we have found that both approaches yield effective results, with no significant difference in performance between them. This suggests that average pooling can retain the multi-level categorical information, similar to concat-and-project. Below we provide the experimental results on the MIND dataset for CTR prediction.

\section{Time Efficiency}

Here, we study the impact of integrating our \model{} module on model efficiency, especially with large-scale datasets, we report both the training time (per epoch) and inference time in various recommendation scenarios. The results indicate that our \model{} module introduces only a marginal increase in computational cost, which is generally negligible in the majority of cases.

\begin{table}[]
\centering
\renewcommand\arraystretch{1.15}
\begin{tabular}{c|cccc}
\toprule
 & \textbf{Training} & \textbf{Training} & \textbf{Inference} & \textbf{Inference} \\
\midrule
\textbf{Base model} & Ori. & \model{} & Ori. & \model{} \\
\midrule
\multicolumn{5}{c}{\textit{CF Scenario (Kindle Dataset)}} \\
\midrule
\textbf{LightGCN} & 14s & 15s & 4s & 4s \\
\textbf{CFKG} & 41s & 44s & 2s & 2s \\
\midrule
\multicolumn{5}{c}{\textit{CTR Scenario (MIND dataset)}} \\
\midrule
\textbf{DCN} & 35s & 37s & 2s & 2s \\
\textbf{FinalMLP} & 29s & 31s & 2s & 2s \\
\midrule
\multicolumn{5}{c}{\textit{List Completion Scenario (Zhihu Dataset)}} \\
\midrule
\textbf{FANS} & 48s & 49s & 69s & 73s \\
\textbf{Bert4Rec} & 38s & 39s & 275s & 287s \\
\bottomrule
\end{tabular}
\caption{Training and inference time comparisons between the original base models and \model{}-enhanced ones. ``Ori.'' represents the original model.}
\end{table}

%% file: tables/statistics.tex
\begin{table*}[]

\centering
\renewcommand\arraystretch{1.2}

\setlength\tabcolsep{2pt}

\resizebox{.9\linewidth}{!}{

% \begin{tabular}{c|ccc|cc|cc}
% \toprule
% & \multicolumn{3}{c|}{\textbf{List Completion}}     & \multicolumn{2}{c|}{\textbf{CTR}} & \multicolumn{2}{c}{\textbf{CF}} \\
% \cmidrule(lr){2-4} \cmidrule(lr){5-6} \cmidrule(lr){7-8}
% \textbf{Datasets} & \textbf{Zhihu} & \textbf{Spotify} & \textbf{Goodreads} & \textbf{MIND} & \textbf{MovieLens} & \textbf{Toys} & \textbf{Kindle} \\
% \midrule
% \textbf{$\#$Users}      & 18,704      & 72,152      & 15,426      & 94,057    & 943     & 19,413    & 68,224    \\
% \textbf{$\#$Items}      & 36,005      & 104,695     & 47,877      & 65,238    & 1,682   & 11,925    & 61,935    \\
% \textbf{$\#$Interactions} & 927,781        & 6,809,820        & 1,589,480          & 1,756,555     & 52,480             & 623,023       & 2,664,795       \\
% \textbf{Items per list} & 49.59       & 94.38       & 103.04      & -         & -       & -         & -         \\
% \textbf{List Range}     & $10\sim200$ & $20\sim300$ & $20\sim300$ & -         & -       & -         & -         \\
% \textbf{Samples}        & -           & -           & -           & 9,993,270 & 69,881  & 1,246,064 & 5,329,590 \\
% \textbf{Density}        & 0.138\%     & 0.089\%     & 0.215\%     & 0.163\%   & 4.406\% & 0.538\%   & 0.136\%  \\
% \bottomrule
% \end{tabular}

% \begin{table}[]
\begin{tabular}{c|ccc|cc|ccccc}
\toprule
 & \multicolumn{3}{c|}{\textbf{List Completion}} & \multicolumn{2}{c|}{\textbf{CTR}} & \multicolumn{5}{c}{\textbf{CF}} \\
\midrule
\textbf{Datasets} & \textbf{Zhihu} & \textbf{Spotify} & \textbf{Goodreads} & \textbf{MIND} & \textbf{MovieLens-100K} & \textbf{Toys} & \textbf{Kindle} & \textbf{Phones} & \textbf{Geocery} & \textbf{MovieLens-1M} \\
\midrule
\textbf{$\#$Users} & 18,704 & 72,152 & 15,426 & 94,057 & 943 & 19,413 & 68,224 & 27,880 & 14,682 & 6,041 \\
\textbf{$\#$Items} & 36,005 & 104,695 & 47,877 & 65,238 & 1,682 & 11,925 & 61,935 & 10,430 & 8,714 & 3,707 \\
\textbf{$\#$Interactions} & 927,781 & 6,809,820 & 1,589,480 & 1,756,555 & 52,480 & 623,023 & 2,664,795 & 194,439 & 151,254 & 1,000,209 \\
\textbf{Items per list} & 49.59 & 94.38 & 103.04 & - & - & - & - & - & - & - \\
\textbf{List Range} & $10\sim200$ & $20\sim300$ & $20\sim300$ & - & - & - & - & - & - & - \\
\textbf{Samples} & - & - & - & 9,993,270 & 69,881 & 1,246,064 & 5,329,590 & 333,120 & 273,146 & 1,988,338 \\
\textbf{Density} & 0.138\% & 0.089\% & 0.215\% & 0.163\% & 4.406\% & 0.538\% & 0.136\% & 0.11\% & 0.21\% & 8.88\% \\
\bottomrule
\end{tabular}
% \end{table}

}

\caption{\label{tab:statistics}Dataset statistics. Density is defined as the ratio of the number of samples to the number of all possible interactions.}

\end{table*}

%% file: CAGE.bbl

\begin{thebibliography}{56}

%%% the Web address.

\ifx \showCODEN    \undefined \def \showCODEN     #1{\unskip}     \fi
\ifx \showDOI      \undefined \def \showDOI       #1{#1}\fi
\ifx \showISBNx    \undefined \def \showISBNx     #1{\unskip}     \fi
\ifx \showISBNxiii \undefined \def \showISBNxiii  #1{\unskip}     \fi
\ifx \showISSN     \undefined \def \showISSN      #1{\unskip}     \fi
\ifx \showLCCN     \undefined \def \showLCCN      #1{\unskip}     \fi
\ifx \shownote     \undefined \def \shownote      #1{#1}          \fi
\ifx \showarticletitle \undefined \def \showarticletitle #1{#1}   \fi
\ifx \showURL      \undefined \def \showURL       {\relax}        \fi
% invisible to TeX
\providecommand\bibfield[2]{#2}
\providecommand\bibinfo[2]{#2}
\providecommand\natexlab[1]{#1}
\providecommand\showeprint[2][]{arXiv:#2}

\bibitem[Babenko and Lempitsky(2014)]%
        {babenko2014additive}
\bibfield{author}{\bibinfo{person}{Artem Babenko} {and} \bibinfo{person}{Victor
  Lempitsky}.} \bibinfo{year}{2014}\natexlab{}.
\newblock \showarticletitle{Additive quantization for extreme vector
  compression}. In \bibinfo{booktitle}{\emph{Proceedings of the IEEE Conference
  on Computer Vision and Pattern Recognition}}. \bibinfo{pages}{931--938}.
\newblock


\bibitem[Barman et~al\mbox{.}(2019)]%
        {barman2019genre}
\bibfield{author}{\bibinfo{person}{Surajit~Das Barman},
  \bibinfo{person}{Mahamudul Hasan}, {and} \bibinfo{person}{Falguni Roy}.}
  \bibinfo{year}{2019}\natexlab{}.
\newblock \showarticletitle{A genre-based item-item collaborative filtering:
  facing the cold-start problem}. In \bibinfo{booktitle}{\emph{Proceedings of
  the 2019 8th international conference on software and computer
  applications}}. \bibinfo{pages}{258--262}.
\newblock


\bibitem[Bengio et~al\mbox{.}(2013)]%
        {bengio2013estimating}
\bibfield{author}{\bibinfo{person}{Yoshua Bengio}, \bibinfo{person}{Nicholas
  L{\'e}onard}, {and} \bibinfo{person}{Aaron Courville}.}
  \bibinfo{year}{2013}\natexlab{}.
\newblock \showarticletitle{Estimating or propagating gradients through
  stochastic neurons for conditional computation}.
\newblock \bibinfo{journal}{\emph{arXiv preprint arXiv:1308.3432}}
  (\bibinfo{year}{2013}).
\newblock


\bibitem[Breese et~al\mbox{.}(2013)]%
        {breese2013empirical}
\bibfield{author}{\bibinfo{person}{John~S Breese}, \bibinfo{person}{David
  Heckerman}, {and} \bibinfo{person}{Carl Kadie}.}
  \bibinfo{year}{2013}\natexlab{}.
\newblock \showarticletitle{Empirical analysis of predictive algorithms for
  collaborative filtering}.
\newblock \bibinfo{journal}{\emph{arXiv preprint arXiv:1301.7363}}
  (\bibinfo{year}{2013}).
\newblock


\bibitem[Buzo et~al\mbox{.}(1980)]%
        {buzo1980speech}
\bibfield{author}{\bibinfo{person}{Andr{\'e}s Buzo}, \bibinfo{person}{A Gray},
  \bibinfo{person}{RM Gray}, {and} \bibinfo{person}{John Markel}.}
  \bibinfo{year}{1980}\natexlab{}.
\newblock \showarticletitle{Speech coding based upon vector quantization}.
\newblock \bibinfo{journal}{\emph{IEEE Transactions on Acoustics, Speech, and
  Signal Processing}} \bibinfo{volume}{28}, \bibinfo{number}{5}
  (\bibinfo{year}{1980}), \bibinfo{pages}{562--574}.
\newblock


\bibitem[Cai et~al\mbox{.}(2021)]%
        {cai2021category}
\bibfield{author}{\bibinfo{person}{Renqin Cai}, \bibinfo{person}{Jibang Wu},
  \bibinfo{person}{Aidan San}, \bibinfo{person}{Chong Wang}, {and}
  \bibinfo{person}{Hongning Wang}.} \bibinfo{year}{2021}\natexlab{}.
\newblock \showarticletitle{Category-aware collaborative sequential
  recommendation}. In \bibinfo{booktitle}{\emph{Proceedings of the 44th
  international ACM SIGIR conference on research and development in information
  retrieval}}. \bibinfo{pages}{388--397}.
\newblock


\bibitem[Chen et~al\mbox{.}(2018)]%
        {chen2018recsys}
\bibfield{author}{\bibinfo{person}{Ching-Wei Chen}, \bibinfo{person}{Paul
  Lamere}, \bibinfo{person}{Markus Schedl}, {and} \bibinfo{person}{Hamed
  Zamani}.} \bibinfo{year}{2018}\natexlab{}.
\newblock \showarticletitle{Recsys Challenge 2018: Automatic Music Playlist
  Continuation}. In \bibinfo{booktitle}{\emph{Proceedings of the 12th ACM
  Conference on Recommender Systems}} (Vancouver, British Columbia, Canada)
  \emph{(\bibinfo{series}{RecSys '18})}. \bibinfo{publisher}{Association for
  Computing Machinery}, \bibinfo{address}{New York, NY, USA},
  \bibinfo{pages}{527–528}.
\newblock
\showISBNx{9781450359016}
\urldef\tempurl%
\url{https://doi.org/10.1145/3240323.3240342}
\showDOI{\tempurl}


\bibitem[Chen et~al\mbox{.}(2012)]%
        {chen2012playlist}
\bibfield{author}{\bibinfo{person}{Shuo Chen}, \bibinfo{person}{Josh~L. Moore},
  \bibinfo{person}{Douglas Turnbull}, {and} \bibinfo{person}{Thorsten
  Joachims}.} \bibinfo{year}{2012}\natexlab{}.
\newblock \showarticletitle{Playlist Prediction via Metric Embedding}. In
  \bibinfo{booktitle}{\emph{Proceedings of the 18th ACM SIGKDD International
  Conference on Knowledge Discovery and Data Mining}} (Beijing, China)
  \emph{(\bibinfo{series}{KDD '12})}. \bibinfo{publisher}{Association for
  Computing Machinery}, \bibinfo{address}{New York, NY, USA},
  \bibinfo{pages}{714–722}.
\newblock
\showISBNx{9781450314626}
\urldef\tempurl%
\url{https://doi.org/10.1145/2339530.2339643}
\showDOI{\tempurl}


\bibitem[Cheng et~al\mbox{.}(2016)]%
        {cheng2016wide}
\bibfield{author}{\bibinfo{person}{Heng-Tze Cheng}, \bibinfo{person}{Levent
  Koc}, \bibinfo{person}{Jeremiah Harmsen}, \bibinfo{person}{Tal Shaked},
  \bibinfo{person}{Tushar Chandra}, \bibinfo{person}{Hrishi Aradhye},
  \bibinfo{person}{Glen Anderson}, \bibinfo{person}{Greg Corrado},
  \bibinfo{person}{Wei Chai}, \bibinfo{person}{Mustafa Ispir}, {et~al\mbox{.}}}
  \bibinfo{year}{2016}\natexlab{}.
\newblock \showarticletitle{Wide \& deep learning for recommender systems}. In
  \bibinfo{booktitle}{\emph{Proceedings of the 1st workshop on deep learning
  for recommender systems}}. \bibinfo{pages}{7--10}.
\newblock


\bibitem[Gatzioura et~al\mbox{.}(2019)]%
        {gatzioura2019hybrid}
\bibfield{author}{\bibinfo{person}{Anna Gatzioura}, \bibinfo{person}{Jo{\~a}o
  Vinagre}, \bibinfo{person}{Al{\'\i}pio~M{\'a}rio Jorge}, {and}
  \bibinfo{person}{Miquel Sanchez-Marre}.} \bibinfo{year}{2019}\natexlab{}.
\newblock \showarticletitle{A Hybrid Recommender System for Improving Automatic
  Playlist Continuation}.
\newblock \bibinfo{journal}{\emph{IEEE Transactions on Knowledge and Data
  Engineering}} \bibinfo{volume}{33}, \bibinfo{number}{5}
  (\bibinfo{year}{2019}), \bibinfo{pages}{1819--1830}.
\newblock


\bibitem[Gogna and Majumdar(2015)]%
        {gogna2015comprehensive}
\bibfield{author}{\bibinfo{person}{Anupriya Gogna} {and}
  \bibinfo{person}{Angshul Majumdar}.} \bibinfo{year}{2015}\natexlab{}.
\newblock \showarticletitle{A comprehensive recommender system model: Improving
  accuracy for both warm and cold start users}.
\newblock \bibinfo{journal}{\emph{IEEE Access}}  \bibinfo{volume}{3}
  (\bibinfo{year}{2015}), \bibinfo{pages}{2803--2813}.
\newblock


\bibitem[Gray(1984)]%
        {gray1984vector}
\bibfield{author}{\bibinfo{person}{Robert Gray}.}
  \bibinfo{year}{1984}\natexlab{}.
\newblock \showarticletitle{Vector quantization}.
\newblock \bibinfo{journal}{\emph{IEEE Assp Magazine}} \bibinfo{volume}{1},
  \bibinfo{number}{2} (\bibinfo{year}{1984}), \bibinfo{pages}{4--29}.
\newblock


\bibitem[Guo et~al\mbox{.}(2018)]%
        {deepfm}
\bibfield{author}{\bibinfo{person}{Huifeng Guo}, \bibinfo{person}{Ruiming
  Tang}, \bibinfo{person}{Yunming Ye}, \bibinfo{person}{Zhenguo Li},
  \bibinfo{person}{Xiuqiang He}, {and} \bibinfo{person}{Zhenhua Dong}.}
  \bibinfo{year}{2018}\natexlab{}.
\newblock \showarticletitle{Deepfm: An end-to-end wide \& deep learning
  framework for CTR prediction}.
\newblock \bibinfo{journal}{\emph{arXiv preprint arXiv:1804.04950}}
  (\bibinfo{year}{2018}).
\newblock


\bibitem[Harper and Konstan(2015)]%
        {movielens}
\bibfield{author}{\bibinfo{person}{F~Maxwell Harper} {and}
  \bibinfo{person}{Joseph~A Konstan}.} \bibinfo{year}{2015}\natexlab{}.
\newblock \showarticletitle{The movielens datasets: History and context}.
\newblock \bibinfo{journal}{\emph{Acm transactions on interactive intelligent
  systems (tiis)}} \bibinfo{volume}{5}, \bibinfo{number}{4}
  (\bibinfo{year}{2015}), \bibinfo{pages}{1--19}.
\newblock


\bibitem[He et~al\mbox{.}(2020a)]%
        {lightgcn}
\bibfield{author}{\bibinfo{person}{Xiangnan He}, \bibinfo{person}{Kuan Deng},
  \bibinfo{person}{Xiang Wang}, \bibinfo{person}{Yan Li},
  \bibinfo{person}{Yongdong Zhang}, {and} \bibinfo{person}{Meng Wang}.}
  \bibinfo{year}{2020}\natexlab{a}.
\newblock \showarticletitle{Lightgcn: Simplifying and powering graph
  convolution network for recommendation}. In
  \bibinfo{booktitle}{\emph{Proceedings of the 43rd International ACM SIGIR
  conference on research and development in Information Retrieval}}.
  \bibinfo{pages}{639--648}.
\newblock


\bibitem[He et~al\mbox{.}(2017)]%
        {neumf}
\bibfield{author}{\bibinfo{person}{Xiangnan He}, \bibinfo{person}{Lizi Liao},
  \bibinfo{person}{Hanwang Zhang}, \bibinfo{person}{Liqiang Nie},
  \bibinfo{person}{Xia Hu}, {and} \bibinfo{person}{Tat-Seng Chua}.}
  \bibinfo{year}{2017}\natexlab{}.
\newblock \showarticletitle{Neural collaborative filtering}. In
  \bibinfo{booktitle}{\emph{Proceedings of the 26th international conference on
  world wide web}}. \bibinfo{pages}{173--182}.
\newblock


\bibitem[He et~al\mbox{.}(2020b)]%
        {car}
\bibfield{author}{\bibinfo{person}{Yun He}, \bibinfo{person}{Yin Zhang},
  \bibinfo{person}{Weiwen Liu}, {and} \bibinfo{person}{James Caverlee}.}
  \bibinfo{year}{2020}\natexlab{b}.
\newblock \showarticletitle{Consistency-Aware Recommendation for User-Generated
  Item List Continuation}. In \bibinfo{booktitle}{\emph{Proceedings of the 13th
  International Conference on Web Search and Data Mining}} (Houston, TX, USA)
  \emph{(\bibinfo{series}{WSDM '20})}. \bibinfo{publisher}{Association for
  Computing Machinery}, \bibinfo{address}{New York, NY, USA},
  \bibinfo{pages}{250–258}.
\newblock
\showISBNx{9781450368223}
\urldef\tempurl%
\url{https://doi.org/10.1145/3336191.3371776}
\showDOI{\tempurl}


\bibitem[Hidasi et~al\mbox{.}(2016)]%
        {gru4rec}
\bibfield{author}{\bibinfo{person}{Bal{\'{a}}zs Hidasi},
  \bibinfo{person}{Alexandros Karatzoglou}, \bibinfo{person}{Linas Baltrunas},
  {and} \bibinfo{person}{Domonkos Tikk}.} \bibinfo{year}{2016}\natexlab{}.
\newblock \showarticletitle{Session-based Recommendations with Recurrent Neural
  Networks}. In \bibinfo{booktitle}{\emph{4th International Conference on
  Learning Representations, {ICLR} 2016, San Juan, Puerto Rico, May 2-4, 2016,
  Conference Track Proceedings}}, \bibfield{editor}{\bibinfo{person}{Yoshua
  Bengio} {and} \bibinfo{person}{Yann LeCun}} (Eds.).
\newblock


\bibitem[Hou et~al\mbox{.}(2023)]%
        {hou2023learning}
\bibfield{author}{\bibinfo{person}{Yupeng Hou}, \bibinfo{person}{Zhankui He},
  \bibinfo{person}{Julian McAuley}, {and} \bibinfo{person}{Wayne~Xin Zhao}.}
  \bibinfo{year}{2023}\natexlab{}.
\newblock \showarticletitle{Learning vector-quantized item representation for
  transferable sequential recommenders}. In
  \bibinfo{booktitle}{\emph{Proceedings of the ACM Web Conference 2023}}.
  \bibinfo{pages}{1162--1171}.
\newblock


\bibitem[Huang et~al\mbox{.}(2019)]%
        {fibinet}
\bibfield{author}{\bibinfo{person}{Tongwen Huang}, \bibinfo{person}{Zhiqi
  Zhang}, {and} \bibinfo{person}{Junlin Zhang}.}
  \bibinfo{year}{2019}\natexlab{}.
\newblock \showarticletitle{FiBiNET: combining feature importance and bilinear
  feature interaction for click-through rate prediction}. In
  \bibinfo{booktitle}{\emph{Proceedings of the 13th ACM Conference on
  Recommender Systems}}. \bibinfo{pages}{169--177}.
\newblock


\bibitem[J{\"a}rvelin and Kek{\"a}l{\"a}inen(2002)]%
        {ndcg}
\bibfield{author}{\bibinfo{person}{Kalervo J{\"a}rvelin} {and}
  \bibinfo{person}{Jaana Kek{\"a}l{\"a}inen}.} \bibinfo{year}{2002}\natexlab{}.
\newblock \showarticletitle{Cumulated gain-based evaluation of IR techniques}.
\newblock \bibinfo{journal}{\emph{ACM Transactions on Information Systems
  (TOIS)}} \bibinfo{volume}{20}, \bibinfo{number}{4} (\bibinfo{year}{2002}),
  \bibinfo{pages}{422--446}.
\newblock


\bibitem[Juang and Gray(1982)]%
        {juang1982multiple}
\bibfield{author}{\bibinfo{person}{Biing-Hwang Juang} {and} \bibinfo{person}{A
  Gray}.} \bibinfo{year}{1982}\natexlab{}.
\newblock \showarticletitle{Multiple stage vector quantization for speech
  coding}. In \bibinfo{booktitle}{\emph{ICASSP'82. IEEE International
  Conference on Acoustics, Speech, and Signal Processing}},
  Vol.~\bibinfo{volume}{7}. IEEE, \bibinfo{pages}{597--600}.
\newblock


\bibitem[Kang and McAuley(2018)]%
        {sasrec}
\bibfield{author}{\bibinfo{person}{Wang-Cheng Kang} {and}
  \bibinfo{person}{Julian McAuley}.} \bibinfo{year}{2018}\natexlab{}.
\newblock \showarticletitle{Self-Attentive Sequential Recommendation}. In
  \bibinfo{booktitle}{\emph{2018 IEEE International Conference on Data Mining
  (ICDM)}}. \bibinfo{publisher}{IEEE}, \bibinfo{pages}{197--206}.
\newblock
\urldef\tempurl%
\url{https://doi.org/10.1109/ICDM.2018.00035}
\showDOI{\tempurl}


\bibitem[Ko et~al\mbox{.}(2021)]%
        {ko2021mascot}
\bibfield{author}{\bibinfo{person}{Yunyong Ko}, \bibinfo{person}{Jae-Seo Yu},
  \bibinfo{person}{Hong-Kyun Bae}, \bibinfo{person}{Yongjun Park},
  \bibinfo{person}{Dongwon Lee}, {and} \bibinfo{person}{Sang-Wook Kim}.}
  \bibinfo{year}{2021}\natexlab{}.
\newblock \showarticletitle{MASCOT: A Quantization Framework for Efficient
  Matrix Factorization in Recommender Systems}. In
  \bibinfo{booktitle}{\emph{2021 IEEE International Conference on Data Mining
  (ICDM)}}. IEEE, \bibinfo{pages}{290--299}.
\newblock


\bibitem[Koren(2009)]%
        {koren2009bellkor}
\bibfield{author}{\bibinfo{person}{Yehuda Koren}.}
  \bibinfo{year}{2009}\natexlab{}.
\newblock \showarticletitle{The bellkor solution to the netflix grand prize}.
\newblock \bibinfo{journal}{\emph{Netflix prize documentation}}
  \bibinfo{volume}{81}, \bibinfo{number}{2009} (\bibinfo{year}{2009}),
  \bibinfo{pages}{1--10}.
\newblock


\bibitem[Koren et~al\mbox{.}(2009)]%
        {koren2009matrix}
\bibfield{author}{\bibinfo{person}{Yehuda Koren}, \bibinfo{person}{Robert
  Bell}, {and} \bibinfo{person}{Chris Volinsky}.}
  \bibinfo{year}{2009}\natexlab{}.
\newblock \showarticletitle{Matrix factorization techniques for recommender
  systems}.
\newblock \bibinfo{journal}{\emph{Computer}} \bibinfo{volume}{42},
  \bibinfo{number}{8} (\bibinfo{year}{2009}), \bibinfo{pages}{30--37}.
\newblock


\bibitem[Lemire and Maclachlan(2005)]%
        {lemire2005slope}
\bibfield{author}{\bibinfo{person}{Daniel Lemire} {and} \bibinfo{person}{Anna
  Maclachlan}.} \bibinfo{year}{2005}\natexlab{}.
\newblock \showarticletitle{Slope one predictors for online rating-based
  collaborative filtering}. In \bibinfo{booktitle}{\emph{Proceedings of the
  2005 SIAM International Conference on Data Mining}}. SIAM,
  \bibinfo{pages}{471--475}.
\newblock


\bibitem[Lian et~al\mbox{.}(2020)]%
        {lian2020product}
\bibfield{author}{\bibinfo{person}{Defu Lian}, \bibinfo{person}{Xing Xie},
  \bibinfo{person}{Enhong Chen}, {and} \bibinfo{person}{Hui Xiong}.}
  \bibinfo{year}{2020}\natexlab{}.
\newblock \showarticletitle{Product quantized collaborative filtering}.
\newblock \bibinfo{journal}{\emph{IEEE Transactions on Knowledge and Data
  Engineering}} \bibinfo{volume}{33}, \bibinfo{number}{9}
  (\bibinfo{year}{2020}), \bibinfo{pages}{3284--3296}.
\newblock


\bibitem[Liu et~al\mbox{.}(2022)]%
        {liu2022prec}
\bibfield{author}{\bibinfo{person}{Qijiong Liu}, \bibinfo{person}{Jieming Zhu},
  \bibinfo{person}{Quanyu Dai}, {and} \bibinfo{person}{Xiaoming Wu}.}
  \bibinfo{year}{2022}\natexlab{}.
\newblock \showarticletitle{Boosting Deep CTR Prediction with a Plug-and-Play
  Pre-trainer for News Recommendation}. In
  \bibinfo{booktitle}{\emph{Proceedings of the 29th International Conference on
  Computational Linguistics}}. \bibinfo{pages}{2823--2833}.
\newblock


\bibitem[Liu et~al\mbox{.}(2023)]%
        {fans}
\bibfield{author}{\bibinfo{person}{Qijiong Liu}, \bibinfo{person}{Jieming Zhu},
  \bibinfo{person}{Jiahao Wu}, \bibinfo{person}{Tiandeng Wu},
  \bibinfo{person}{Zhenhua Dong}, {and} \bibinfo{person}{Xiao-Ming Wu}.}
  \bibinfo{year}{2023}\natexlab{}.
\newblock \showarticletitle{FANS: Fast Non-Autoregressive Sequence Generation
  for Item List Continuation}. In \bibinfo{booktitle}{\emph{Proceedings of the
  ACM Web Conference 2023}}. \bibinfo{pages}{3309--3318}.
\newblock


\bibitem[Mao et~al\mbox{.}(2023)]%
        {mao2023finalmlp}
\bibfield{author}{\bibinfo{person}{Kelong Mao}, \bibinfo{person}{Jieming Zhu},
  \bibinfo{person}{Liangcai Su}, \bibinfo{person}{Guohao Cai},
  \bibinfo{person}{Yuru Li}, {and} \bibinfo{person}{Zhenhua Dong}.}
  \bibinfo{year}{2023}\natexlab{}.
\newblock \showarticletitle{FinalMLP: an enhanced two-stream MLP model for CTR
  prediction}. In \bibinfo{booktitle}{\emph{Proceedings of the AAAI Conference
  on Artificial Intelligence}}, Vol.~\bibinfo{volume}{37}.
  \bibinfo{pages}{4552--4560}.
\newblock


\bibitem[McFee and Lanckriet(2011)]%
        {mcfee2011natural}
\bibfield{author}{\bibinfo{person}{Brian McFee} {and} \bibinfo{person}{Gert~RG
  Lanckriet}.} \bibinfo{year}{2011}\natexlab{}.
\newblock \showarticletitle{The Natural Language of Playlists.}. In
  \bibinfo{booktitle}{\emph{The International Society for Music Information
  Retrieval (ISMIR)}}, Vol.~\bibinfo{volume}{11}. \bibinfo{pages}{537--541}.
\newblock


\bibitem[McFee and Lanckriet(2012)]%
        {mcfee2012hypergraph}
\bibfield{author}{\bibinfo{person}{Brian McFee} {and} \bibinfo{person}{Gert~RG
  Lanckriet}.} \bibinfo{year}{2012}\natexlab{}.
\newblock \showarticletitle{Hypergraph Models of Playlist Dialects.}. In
  \bibinfo{booktitle}{\emph{The International Society for Music Information
  Retrieval (ISMIR)}}, Vol.~\bibinfo{volume}{12}. \bibinfo{pages}{343--348}.
\newblock


\bibitem[Moreira et~al\mbox{.}(2019)]%
        {moreira2019importance}
\bibfield{author}{\bibinfo{person}{Gabriel de Souza~P Moreira},
  \bibinfo{person}{Dietmar Jannach}, {and} \bibinfo{person}{Adilson~Marques da
  Cunha}.} \bibinfo{year}{2019}\natexlab{}.
\newblock \showarticletitle{On the importance of news content representation in
  hybrid neural session-based recommender systems}.
\newblock \bibinfo{journal}{\emph{arXiv preprint arXiv:1907.07629}}
  (\bibinfo{year}{2019}).
\newblock


\bibitem[Pan et~al\mbox{.}(2021)]%
        {pan2021click}
\bibfield{author}{\bibinfo{person}{Yujie Pan}, \bibinfo{person}{Jiangchao Yao},
  \bibinfo{person}{Bo Han}, \bibinfo{person}{Kunyang Jia}, \bibinfo{person}{Ya
  Zhang}, {and} \bibinfo{person}{Hongxia Yang}.}
  \bibinfo{year}{2021}\natexlab{}.
\newblock \showarticletitle{Click-through rate prediction with auto-quantized
  contrastive learning}.
\newblock \bibinfo{journal}{\emph{arXiv preprint arXiv:2109.13921}}
  (\bibinfo{year}{2021}).
\newblock


\bibitem[Qu et~al\mbox{.}(2016)]%
        {pnn}
\bibfield{author}{\bibinfo{person}{Yanru Qu}, \bibinfo{person}{Han Cai},
  \bibinfo{person}{Kan Ren}, \bibinfo{person}{Weinan Zhang},
  \bibinfo{person}{Yong Yu}, \bibinfo{person}{Ying Wen}, {and}
  \bibinfo{person}{Jun Wang}.} \bibinfo{year}{2016}\natexlab{}.
\newblock \showarticletitle{Product-based neural networks for user response
  prediction}. In \bibinfo{booktitle}{\emph{2016 IEEE 16th international
  conference on data mining (ICDM)}}. IEEE, \bibinfo{pages}{1149--1154}.
\newblock


\bibitem[Rajput et~al\mbox{.}(2024)]%
        {rajput2023recommender}
\bibfield{author}{\bibinfo{person}{Shashank Rajput}, \bibinfo{person}{Nikhil
  Mehta}, \bibinfo{person}{Anima Singh}, \bibinfo{person}{Raghunandan
  Hulikal~Keshavan}, \bibinfo{person}{Trung Vu}, \bibinfo{person}{Lukasz
  Heldt}, \bibinfo{person}{Lichan Hong}, \bibinfo{person}{Yi Tay},
  \bibinfo{person}{Vinh Tran}, \bibinfo{person}{Jonah Samost}, {et~al\mbox{.}}}
  \bibinfo{year}{2024}\natexlab{}.
\newblock \showarticletitle{Recommender systems with generative retrieval}.
\newblock \bibinfo{journal}{\emph{Advances in Neural Information Processing
  Systems}}  \bibinfo{volume}{36} (\bibinfo{year}{2024}).
\newblock


\bibitem[Razavi et~al\mbox{.}(2019)]%
        {vqvae2}
\bibfield{author}{\bibinfo{person}{Ali Razavi}, \bibinfo{person}{Aaron Van~den
  Oord}, {and} \bibinfo{person}{Oriol Vinyals}.}
  \bibinfo{year}{2019}\natexlab{}.
\newblock \showarticletitle{Generating diverse high-fidelity images with
  vq-vae-2}.
\newblock \bibinfo{journal}{\emph{Advances in neural information processing
  systems}}  \bibinfo{volume}{32} (\bibinfo{year}{2019}).
\newblock


\bibitem[Rendle et~al\mbox{.}(2012)]%
        {bpr}
\bibfield{author}{\bibinfo{person}{Steffen Rendle}, \bibinfo{person}{Christoph
  Freudenthaler}, \bibinfo{person}{Zeno Gantner}, {and} \bibinfo{person}{Lars
  Schmidt-Thieme}.} \bibinfo{year}{2012}\natexlab{}.
\newblock \showarticletitle{BPR: Bayesian personalized ranking from implicit
  feedback}.
\newblock \bibinfo{journal}{\emph{arXiv preprint arXiv:1205.2618}}
  (\bibinfo{year}{2012}).
\newblock


\bibitem[Sarwar et~al\mbox{.}(2001)]%
        {itemcf}
\bibfield{author}{\bibinfo{person}{Badrul Sarwar}, \bibinfo{person}{George
  Karypis}, \bibinfo{person}{Joseph Konstan}, {and} \bibinfo{person}{John
  Riedl}.} \bibinfo{year}{2001}\natexlab{}.
\newblock \showarticletitle{Item-based collaborative filtering recommendation
  algorithms}. In \bibinfo{booktitle}{\emph{Proceedings of the 10th
  international conference on World Wide Web}}. \bibinfo{pages}{285--295}.
\newblock


\bibitem[Shi et~al\mbox{.}(2019)]%
        {shi2019deep}
\bibfield{author}{\bibinfo{person}{Chuan Shi}, \bibinfo{person}{Xiaotian Han},
  \bibinfo{person}{Li Song}, \bibinfo{person}{Xiao Wang},
  \bibinfo{person}{Senzhang Wang}, \bibinfo{person}{Junping Du}, {and}
  \bibinfo{person}{S~Yu Philip}.} \bibinfo{year}{2019}\natexlab{}.
\newblock \showarticletitle{Deep collaborative filtering with multi-aspect
  information in heterogeneous networks}.
\newblock \bibinfo{journal}{\emph{IEEE transactions on knowledge and data
  engineering}} \bibinfo{volume}{33}, \bibinfo{number}{4}
  (\bibinfo{year}{2019}), \bibinfo{pages}{1413--1425}.
\newblock


\bibitem[Sun et~al\mbox{.}(2019)]%
        {bert4rec}
\bibfield{author}{\bibinfo{person}{Fei Sun}, \bibinfo{person}{Jun Liu},
  \bibinfo{person}{Jian Wu}, \bibinfo{person}{Changhua Pei},
  \bibinfo{person}{Xiao Lin}, \bibinfo{person}{Wenwu Ou}, {and}
  \bibinfo{person}{Peng Jiang}.} \bibinfo{year}{2019}\natexlab{}.
\newblock \showarticletitle{BERT4Rec: Sequential Recommendation with
  Bidirectional Encoder Representations from Transformer}. In
  \bibinfo{booktitle}{\emph{Proceedings of the 28th ACM International
  Conference on Information and Knowledge Management}} (Beijing, China)
  \emph{(\bibinfo{series}{CIKM '19})}. \bibinfo{publisher}{Association for
  Computing Machinery}, \bibinfo{address}{New York, NY, USA},
  \bibinfo{pages}{1441–1450}.
\newblock
\showISBNx{9781450369763}
\urldef\tempurl%
\url{https://doi.org/10.1145/3357384.3357895}
\showDOI{\tempurl}


\bibitem[Tang and Wang(2018)]%
        {caser}
\bibfield{author}{\bibinfo{person}{Jiaxi Tang} {and} \bibinfo{person}{Ke
  Wang}.} \bibinfo{year}{2018}\natexlab{}.
\newblock \showarticletitle{Personalized Top-N Sequential Recommendation via
  Convolutional Sequence Embedding} \emph{(\bibinfo{series}{WSDM '18})}.
  \bibinfo{publisher}{Association for Computing Machinery},
  \bibinfo{address}{New York, NY, USA}, \bibinfo{pages}{565–573}.
\newblock
\showISBNx{9781450355810}
\urldef\tempurl%
\url{https://doi.org/10.1145/3159652.3159656}
\showDOI{\tempurl}


\bibitem[Tran et~al\mbox{.}(2019)]%
        {tran2019adversarial}
\bibfield{author}{\bibinfo{person}{Thanh Tran}, \bibinfo{person}{Renee
  Sweeney}, {and} \bibinfo{person}{Kyumin Lee}.}
  \bibinfo{year}{2019}\natexlab{}.
\newblock \showarticletitle{Adversarial Mahalanobis Distance-Based Attentive
  Song Recommender for Automatic Playlist Continuation}. In
  \bibinfo{booktitle}{\emph{Proceedings of the 42nd International ACM SIGIR
  Conference on Research and Development in Information Retrieval}} (Paris,
  France) \emph{(\bibinfo{series}{SIGIR'19})}. \bibinfo{publisher}{Association
  for Computing Machinery}, \bibinfo{address}{New York, NY, USA},
  \bibinfo{pages}{245–254}.
\newblock
\showISBNx{9781450361729}
\urldef\tempurl%
\url{https://doi.org/10.1145/3331184.3331234}
\showDOI{\tempurl}


\bibitem[Van~Balen and Levy(2019)]%
        {van2019pq}
\bibfield{author}{\bibinfo{person}{Jan Van~Balen} {and} \bibinfo{person}{Mark
  Levy}.} \bibinfo{year}{2019}\natexlab{}.
\newblock \showarticletitle{PQ-VAE: Efficient Recommendation Using Quantized
  Embeddings.}. In \bibinfo{booktitle}{\emph{RecSys (Late-Breaking Results)}}.
  \bibinfo{pages}{46--50}.
\newblock


\bibitem[Van Den~Oord et~al\mbox{.}(2017)]%
        {vqvae}
\bibfield{author}{\bibinfo{person}{Aaron Van Den~Oord}, \bibinfo{person}{Oriol
  Vinyals}, {et~al\mbox{.}}} \bibinfo{year}{2017}\natexlab{}.
\newblock \showarticletitle{Neural discrete representation learning}.
\newblock \bibinfo{journal}{\emph{Advances in neural information processing
  systems}}  \bibinfo{volume}{30} (\bibinfo{year}{2017}).
\newblock


\bibitem[Volkovs et~al\mbox{.}(2018)]%
        {volkovs2018two}
\bibfield{author}{\bibinfo{person}{Maksims Volkovs}, \bibinfo{person}{Himanshu
  Rai}, \bibinfo{person}{Zhaoyue Cheng}, \bibinfo{person}{Ga Wu},
  \bibinfo{person}{Yichao Lu}, {and} \bibinfo{person}{Scott Sanner}.}
  \bibinfo{year}{2018}\natexlab{}.
\newblock \showarticletitle{Two-Stage Model for Automatic Playlist Continuation
  at Scale}. In \bibinfo{booktitle}{\emph{Proceedings of the ACM Recommender
  Systems Challenge 2018}} (Vancouver, BC, Canada)
  \emph{(\bibinfo{series}{RecSys Challenge '18})}.
  \bibinfo{publisher}{Association for Computing Machinery},
  \bibinfo{address}{New York, NY, USA}, Article \bibinfo{articleno}{9},
  \bibinfo{numpages}{6}~pages.
\newblock
\showISBNx{9781450365864}
\urldef\tempurl%
\url{https://doi.org/10.1145/3267471.3267480}
\showDOI{\tempurl}


\bibitem[Wang et~al\mbox{.}(2020)]%
        {wang2020make}
\bibfield{author}{\bibinfo{person}{Chenyang Wang}, \bibinfo{person}{Min Zhang},
  \bibinfo{person}{Weizhi Ma}, \bibinfo{person}{Yiqun Liu}, {and}
  \bibinfo{person}{Shaoping Ma}.} \bibinfo{year}{2020}\natexlab{}.
\newblock \showarticletitle{Make it a chorus: knowledge-and time-aware item
  modeling for sequential recommendation}. In
  \bibinfo{booktitle}{\emph{Proceedings of the 43rd International ACM SIGIR
  Conference on Research and Development in Information Retrieval}}.
  \bibinfo{pages}{109--118}.
\newblock


\bibitem[Wang et~al\mbox{.}(2023)]%
        {wang2023towards}
\bibfield{author}{\bibinfo{person}{Fangye Wang}, \bibinfo{person}{Hansu Gu},
  \bibinfo{person}{Dongsheng Li}, \bibinfo{person}{Tun Lu},
  \bibinfo{person}{Peng Zhang}, {and} \bibinfo{person}{Ning Gu}.}
  \bibinfo{year}{2023}\natexlab{}.
\newblock \showarticletitle{Towards Deeper, Lighter and Interpretable Cross
  Network for CTR Prediction}. In \bibinfo{booktitle}{\emph{Proceedings of the
  32nd ACM International Conference on Information and Knowledge Management}}.
  \bibinfo{pages}{2523--2533}.
\newblock


\bibitem[Wang et~al\mbox{.}(2017)]%
        {dcn}
\bibfield{author}{\bibinfo{person}{Ruoxi Wang}, \bibinfo{person}{Bin Fu},
  \bibinfo{person}{Gang Fu}, {and} \bibinfo{person}{Mingliang Wang}.}
  \bibinfo{year}{2017}\natexlab{}.
\newblock \showarticletitle{Deep \& Cross Network for Ad Click Predictions}. In
  \bibinfo{booktitle}{\emph{Proceedings of the ADKDD'17}} (Halifax, NS, Canada)
  \emph{(\bibinfo{series}{ADKDD'17})}. \bibinfo{publisher}{Association for
  Computing Machinery}, \bibinfo{address}{New York, NY, USA}, Article
  \bibinfo{articleno}{12}, \bibinfo{numpages}{7}~pages.
\newblock
\showISBNx{9781450351942}


\bibitem[Wang et~al\mbox{.}(2021)]%
        {wang2021masknet}
\bibfield{author}{\bibinfo{person}{Zhiqiang Wang}, \bibinfo{person}{Qingyun
  She}, {and} \bibinfo{person}{Junlin Zhang}.} \bibinfo{year}{2021}\natexlab{}.
\newblock \showarticletitle{Masknet: Introducing feature-wise multiplication to
  CTR ranking models by instance-guided mask}.
\newblock \bibinfo{journal}{\emph{arXiv preprint arXiv:2102.07619}}
  (\bibinfo{year}{2021}).
\newblock


\bibitem[Wu et~al\mbox{.}(2020)]%
        {mind}
\bibfield{author}{\bibinfo{person}{Fangzhao Wu}, \bibinfo{person}{Ying Qiao},
  \bibinfo{person}{Jiun-Hung Chen}, \bibinfo{person}{Chuhan Wu},
  \bibinfo{person}{Tao Qi}, \bibinfo{person}{Jianxun Lian},
  \bibinfo{person}{Danyang Liu}, \bibinfo{person}{Xing Xie},
  \bibinfo{person}{Jianfeng Gao}, \bibinfo{person}{Winnie Wu}, {et~al\mbox{.}}}
  \bibinfo{year}{2020}\natexlab{}.
\newblock \showarticletitle{Mind: A large-scale dataset for news
  recommendation}. In \bibinfo{booktitle}{\emph{Proceedings of the 58th Annual
  Meeting of the Association for Computational Linguistics}}.
  \bibinfo{pages}{3597--3606}.
\newblock


\bibitem[Wu and Yu(2019)]%
        {wu2019vector}
\bibfield{author}{\bibinfo{person}{Ze-bin Wu} {and} \bibinfo{person}{Jun-qing
  Yu}.} \bibinfo{year}{2019}\natexlab{}.
\newblock \showarticletitle{Vector quantization: a review}.
\newblock \bibinfo{journal}{\emph{Frontiers of Information Technology \&
  Electronic Engineering}} \bibinfo{volume}{20}, \bibinfo{number}{4}
  (\bibinfo{year}{2019}), \bibinfo{pages}{507--524}.
\newblock


\bibitem[Xia et~al\mbox{.}(2013)]%
        {xia2013joint}
\bibfield{author}{\bibinfo{person}{Yan Xia}, \bibinfo{person}{Kaiming He},
  \bibinfo{person}{Fang Wen}, {and} \bibinfo{person}{Jian Sun}.}
  \bibinfo{year}{2013}\natexlab{}.
\newblock \showarticletitle{Joint inverted indexing}. In
  \bibinfo{booktitle}{\emph{Proceedings of the IEEE International Conference on
  Computer Vision}}. \bibinfo{pages}{3416--3423}.
\newblock


\bibitem[Zhang et~al\mbox{.}(2023)]%
        {zhang2023query}
\bibfield{author}{\bibinfo{person}{Jin Zhang}, \bibinfo{person}{Defu Lian},
  \bibinfo{person}{Haodi Zhang}, \bibinfo{person}{Baoyun Wang}, {and}
  \bibinfo{person}{Enhong Chen}.} \bibinfo{year}{2023}\natexlab{}.
\newblock \showarticletitle{Query-Aware Quantization for Maximum Inner Product
  Search}. In \bibinfo{booktitle}{\emph{Proceedings of the AAAI Conference on
  Artificial Intelligence}}, Vol.~\bibinfo{volume}{37}.
  \bibinfo{pages}{4875--4883}.
\newblock


\bibitem[Zhang et~al\mbox{.}(2018)]%
        {cfkg}
\bibfield{author}{\bibinfo{person}{Yongfeng Zhang}, \bibinfo{person}{Qingyao
  Ai}, \bibinfo{person}{Xu Chen}, {and} \bibinfo{person}{Pengfei Wang}.}
  \bibinfo{year}{2018}\natexlab{}.
\newblock \showarticletitle{Learning over knowledge-base embeddings for
  recommendation}.
\newblock \bibinfo{journal}{\emph{arXiv preprint arXiv:1803.06540}}
  (\bibinfo{year}{2018}).
\newblock


\end{thebibliography}
